\documentclass[journal,10pt]{IEEEtran} % [11pt,draftcls,onecolumn][journal][10pt,onecolum,twoside,compsoc]
\usepackage{graphicx}  %[dvips] 不能加！直接转换pdf文件！
\usepackage{epsfig}
\usepackage{color}
\usepackage[cmex10]{amsmath}
\usepackage{amssymb}

\usepackage{array}  % 特殊字体、符号都有了。
\ifCLASSINFOpdf
\else
\fi
\begin{document}
\title{ A 3D Non-Stationary Wideband Geometry-Based Channel Model for MIMO Vehicle-to-Vehicle Communication Systems }
\author{ Hao Jiang, ~Zaichen Zhang, ~\IEEEmembership{Senior Member,~IEEE}, ~Liang Wu, ~\IEEEmembership{Member,~IEEE}, ~Jian Dang, ~\IEEEmembership{Member,~IEEE}, and ~Guan Gui, ~\IEEEmembership{Senior Member,~IEEE}
\thanks{ This work is supported by national key research and development plan (No. 2016YFB0502202), NSFC projects (61571105, 61501109, 61601119, and 61601120), and scientific research foundation of graduate school of Southeast University (No. YBJJ1655). }
\thanks{ H. Jiang, Z. Zhang, L. Wu, and J. Dang are with the National Mobile Communications Research Laboratory, Southeast University, Nanjing 210096, P. R. China. (e-mails: \{jianghao, zczhang, wuliang, newwanda\}@seu.edu.cn), Corresponding author: zczhang@seu.edu.cn). }
\thanks{ G. Gui is with the Department of Telecommunication and Information Engineering, Nanjing University of Posts and Telecommunications, Nanjing 210003, P. R. China (email: guiguan@njupt.edu.cn). } }
\maketitle
\begin{abstract}

In this paper, we present a three-dimensional (3D) non-wide-sense stationary (non-WSS) wideband geometry-based channel model for vehicle-to-vehicle (V2V) communication environments. We introduce a two-cylinder model to describe moving vehicles as well as multiple confocal semi-ellipsoid models to depict stationary roadside scenarios. The received signal is constructed as a sum of the line-of-sight (LoS), single-, and double-bounced rays with different energies. Accordingly, the proposed channel model is sufficient for depicting a wide variety of V2V environments, such as macro-, micro-, and picocells. The relative movement between the mobile transmitter (MT) and mobile receiver (MR) results in time-variant geometric statistics that make our channel model non-stationary. Using this channel model, the proposed channel statistics, i.e., the time-variant space correlation functions (CFs), frequency CFs, and corresponding Doppler power spectral density (PSD), were studied for different relative moving time instants. The numerical results demonstrate that the proposed 3D non-WSS wideband channel model is practical for characterizing real V2V channels.

\end{abstract}
\begin{IEEEkeywords}

3D channel model, vehicle-to-vehicle communication environments, time-variant space and frequency correlation functions, Doppler power spectral density.

\end{IEEEkeywords}
\IEEEpeerreviewmaketitle

\section{Introduction}

\subsection{Motivation}

Recently, vehicle-to-vehicle (V2V) communications have received widespread applications on account of the rapid development of fifth-generation (5G) wireless communication networks [1]. Unlike conventional fixed-to-mobile (F2M) cellular systems, V2V systems are employed with low-elevation multiple antennas, and the mobile transmitter (MT) and mobile receiver (MR) are both in relative motion. In V2V scenarios, multiple-input and multiple-output (MIMO) technology is becoming increasingly attractive because large-scale antenna elements can be easily mounted on vehicular surfaces. For facilitating the design and analysis of V2V communication systems, the radio propagation characteristics must be designed between the MT and MR [2-4]. Reliable knowledge of the realistic propagation channel models, which provide effective and simple means to approximately express the statistical properties of the V2V channel [5-6].

\subsection{Prior Work}

\hspace*{-0.475cm} \emph{1) Geometry-Based Channels}

To evaluate the performance of MIMO V2V communication systems, accurate channel models are indispensable. Regarding the approach of V2V channel modeling, the models can be categorized as deterministic models (mainly indicates the ray-tracing method) and stochastic models. In particular, stochastic models can be roughly divided into several categories as non-geometrical geometry-based stochastic models (NGBSMs) and regular-shaped geometry-based stochastic models (RS-GBSMs) [9]-[17]. The former is also known as parametric models, which is constructed based on the channel measurements, while the latter is based on the regular geometric shape of scatterers.

In [9], the authors demonstrated that the line-of-sight (LoS) is more likely to be obstructed by buildings and obstacles between the MT and MR. Thus, it is necessary to develop Rayleigh channels to describe the V2V environments. The authors of  [11] introduced an RS-GBSM for V2V scenarios. The author presented a two-ring model to depict the moving scatterers and multiple confocal ellipses to mimic static scatterers. Yuan [12] adopted a two-sphere model to describe moving vehicles as well as multiple confocal elliptic-cylinder models to depict stationary roadside scenarios. In 2014, Zajic [13] proposed a two-cylinder model to depict moving and stationary scatterers in the vicinity of the transmitter and receiver. Accordingly, in [13], the authors stated that the mobility of scatterers significantly affects the Doppler spectrum; therefore, it is important to accurately account for that effect. Furthermore, three-dimensional (3D) RS-GBSMs for macrocell and microcell communication environments were respectively presented in [15] and [16]. However, most of the above RS-GBSMs focus on narrowband channel models, wherein all rays experience a similar propagation delay [17]. This unrealistically describes wireless communication environments. According to the channel measurements between the narrowband and wideband V2V channels in [18], Sen concluded that the channel statistics for different time delays in wideband channels should be addressed. In 2009, the authors in [19] first proposed a wideband RS-GBSM for MIMO V2V Ricean fading channels. However, in [19], the model was shown to be unable to describe the channel statistics for different time delays, which are significant for wideband channels. Based on two measured scenarios in [20], Cheng [21] introduced the concept of high vehicle traffic density (VTD) and low VTD to represent moving vehicles, respectively. That author presented the channel statistics for different time delays (i.e., per-tap channel statistics); nevertheless, the angular spreading of incident waves in an elevation plane in 3D space was ignored.

\hspace*{-0.475cm} \emph{2) Non-Stationary Channels}

Most previous channel models rely on the wide-sense stationary uncorrelated scattering (WSSUS) assumption, which adopts a static channel with constant model parameters, cannot be used to depict the dynamic channel properties. To fill the above gaps, it is thus desirable to re-evaluate the validity of non-wide-sense stationary (non-WSS) channel modeling to describe the vehicular communications between an MT and MR.

In [23] and [24], the authors proposed two-dimensional (2D) geometry-based non-WSS narrowband channel models for T-junction and straight road environments, respectively. Additionally, [25] and [26] presented 2D non-stationary theoretical wideband MIMO Ricean channels for V2V scenarios. However, these channels remain restricted to research in an azimuth plane. Furthermore, Yuan [27] presented a 3D wideband MIMO V2V channel. Nonetheless, that author only focused on two relative special moving directions: the same direction and opposite direction. The authors of [28] presented a wideband MIMO model for V2V channels based on extensive measurements taken in highway and rural environments. In their study, the effects of the mobile discrete scatterers, static discrete scatterers, and diffuse scatterers on the time-variant channel properties were investigated. For the above-mentioned channel models, they did not analyze the mobile properties between the MT and MR, including the relative moving time and moving directions [5]. However, the time-variant space-time and frequency CFs, which are meaningful for the wireless channel, were not studied in detail. Therefore, these models cannot realistically describe V2V communication environments.

\subsection{Main Contributions}

\footnotetext[1]{Actually, the proposed band is also capable of some other V2V environments, such as urban and highway scenarios. For example, note that the proposed band is close to the 5.9 GHz V2V band [21]. However, the difference between 5.4 and 5.9 GHz is 9.3\% ($=$ 0.5/5.4); thus, their propagation channel characteristics do not change significantly. Based on the measurements in [38] and [39], the path loss exponent has a variation of less than 15\% over 1 GHz bandwidth and the delay spread has less than 10\% variation over 8 GHz bandwidth. Here, we could regard these values as an uncertainty of the estimated model parameters at 5.4 GHz, when the goal is to estimate parameter values at 5.9 GHz.}

In this paper, we present a 3D non-stationary wideband semi-ellipsoid model for MIMO V2V Ricean fading channels. The model is operated at 5.4 GHz, with a bandwidth of 50 MHz.\footnotemark[1] Compared with the work in [28], the channel model in this paper is capable of depicting a wide variety of communication environments by adjusting the model parameters. Additionally, our model is time-variant because of the relative motion between the MT and MR. Consequently, we can analyze the proposed channel statistics for more moving directions, rather than some special moving conditions as mentioned in [28]. Furthermore, in the proposed model, the effect of road width on the V2V channel statistics can be investigated. It is important to analyze the proposed channel statistics for different taps and different path delays in non-stationary conditions. This model further corrects the unrealistic assumption widely used in current V2V RS-GBSMs. For example, the authors in [5] adopted the WSS channel to describe the V2V scenarios; the impact of non-stationary on V2V channel statistics was neglected. It is assumed that the azimuth angle of departure (AAoD), elevation angle of departure (EAoD), azimuth angle of arrival (AAoA), and elevation angle of arrival (EAoA) are independent of each other [25]. The major contributions of this paper are outlined as follows:

(1) Based on the two measured scenarios mentioned above in [20], we propose a 3D non-stationary wideband geometric channel model for two different V2V communication environments, i.e., highway scenarios and urban scenarios.

(2) We outline the statistical properties of the proposed V2V channel model for different taps. Important time-variant channel statistics are derived and thoroughly investigated. Specifically, the time-variant space and frequency correlation functions (CFs) and corresponding Doppler power spectral densities (PSDs) are derived for V2V scenarios with different relative moving directions.

(3) The impacts of non-stationarity (i.e., relative moving time and relative moving directions) on time-variant space and frequency CFs are investigated in a comparison with those of the corresponding WSS model and measured results. The results show that the proposed channel model is an excellent approximation of the realistic V2V scenarios.

(4) The geometric path lengths between the MT and MR in a 3D semi-ellipsoid V2V channel model continue to change because the transmit azimuth and elevation angles constantly vary. We thus analyze the proposed statistical properties for different taps and different path delays, which is a different approach than those presented in previous works [12,25,27].

The remainder of this paper is organized as follows. Section II details the proposed theoretical 3D non-stationary wideband MIMO V2V channel model. In Section III, based on the proposed geometric model, the time-variant space CFs, frequency CFs, and corresponding Doppler PSDs are derived. Numerical results and discussions are provided in Section IV. Finally, our conclusions are presented in Section V.

\section{ 3D Geometry-Based V2V Theoretical Channel Model }

In V2V scenarios, the impacts of moving vehicles and roadside environments on the channel statistical properties should be addressed [11,21]. Additionally, the relative movement between the MT and MR makes the V2V channel time-variant. However, the previous channel models have certain limitations in terms of realistically describing the V2V communication environments. For example, the models in [10] and [14] rely on the WSS assumption, which implies that in the time domain, the statistical properties of the channel remain invariant over a short period of time. Thus, the above channel models could not depict the real V2V environments because of the motion between the MT and MR. The authors in [15] and [19] presented the semi-ellipsoid and cylinder models, respectively, to describe the scatterers surrounding the transmitter and receiver. However, in these studies, the effect of the roadside environments on the channel characteristics was not discussed. In [31] and [32], the authors proposed ellipsoid channel models to describe the mobile radio environments. However, the moving vehicles around the MT and MR were not investigated in V2V environments. On the other hand, the authors in [28] performed the channel measurements only with the MT and MR driving in the same direction, and with the MT and MR driving in opposite directions. However, the effect of the arbitrary moving directions on the channel statistics was not investigated. Motivated by the above drawbacks, we have adopted a 3D non-stationary wideband geometric channel model in this paper to depict the actual vehicular communications, as illustrated in Figs. 1 and 2.

\begin{figure}[t]
\begin{center}
\epsfig{file=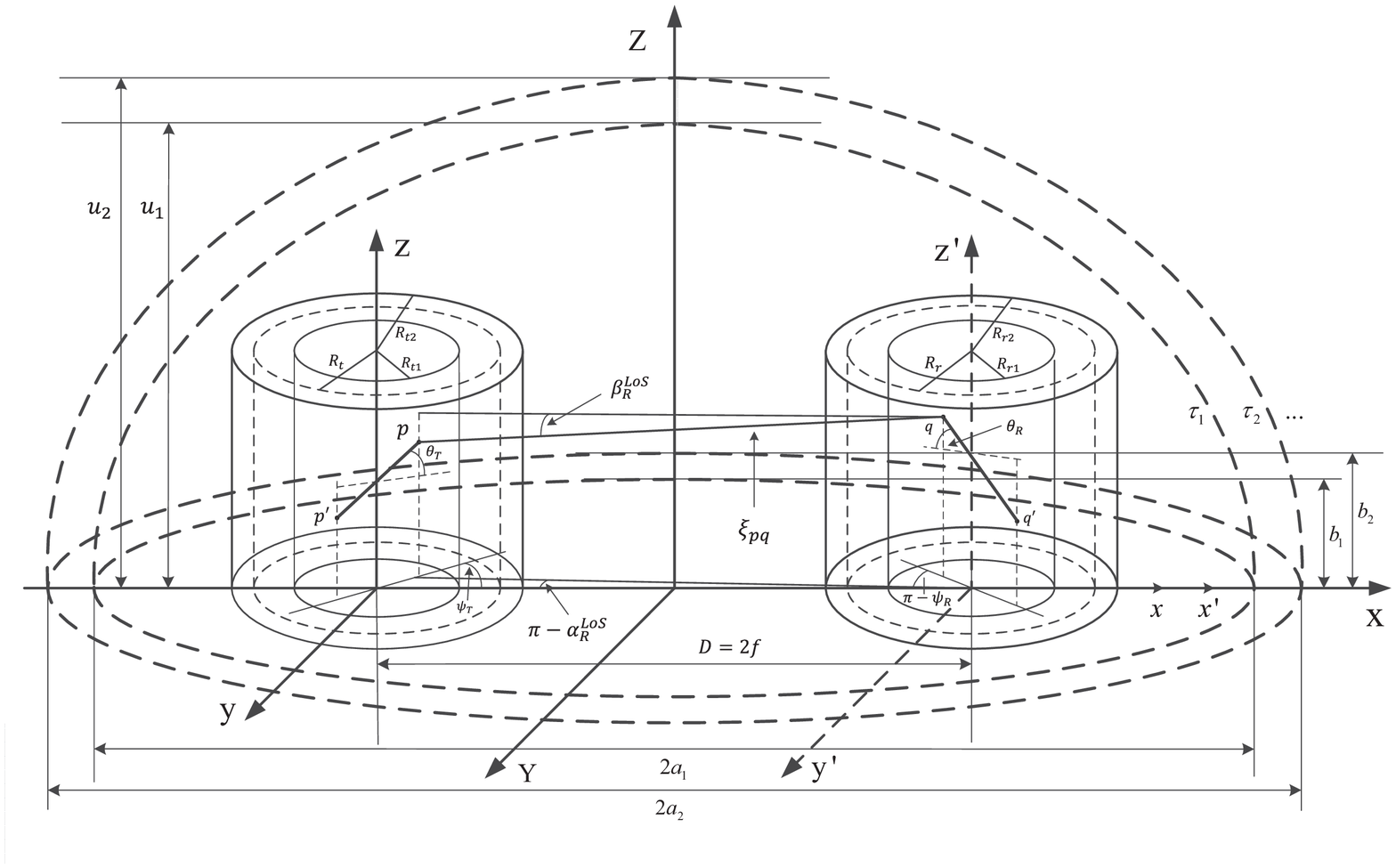,width=9.1cm} \caption{ Proposed 3D wideband MIMO V2V channel model combining the two-cylinder model and multiple confocal semi-ellipsoid models for the line-of-sight (LoS) propagation rays. }
\end{center}
\end{figure}
\begin{figure}[t]
\begin{center}
\epsfig{file=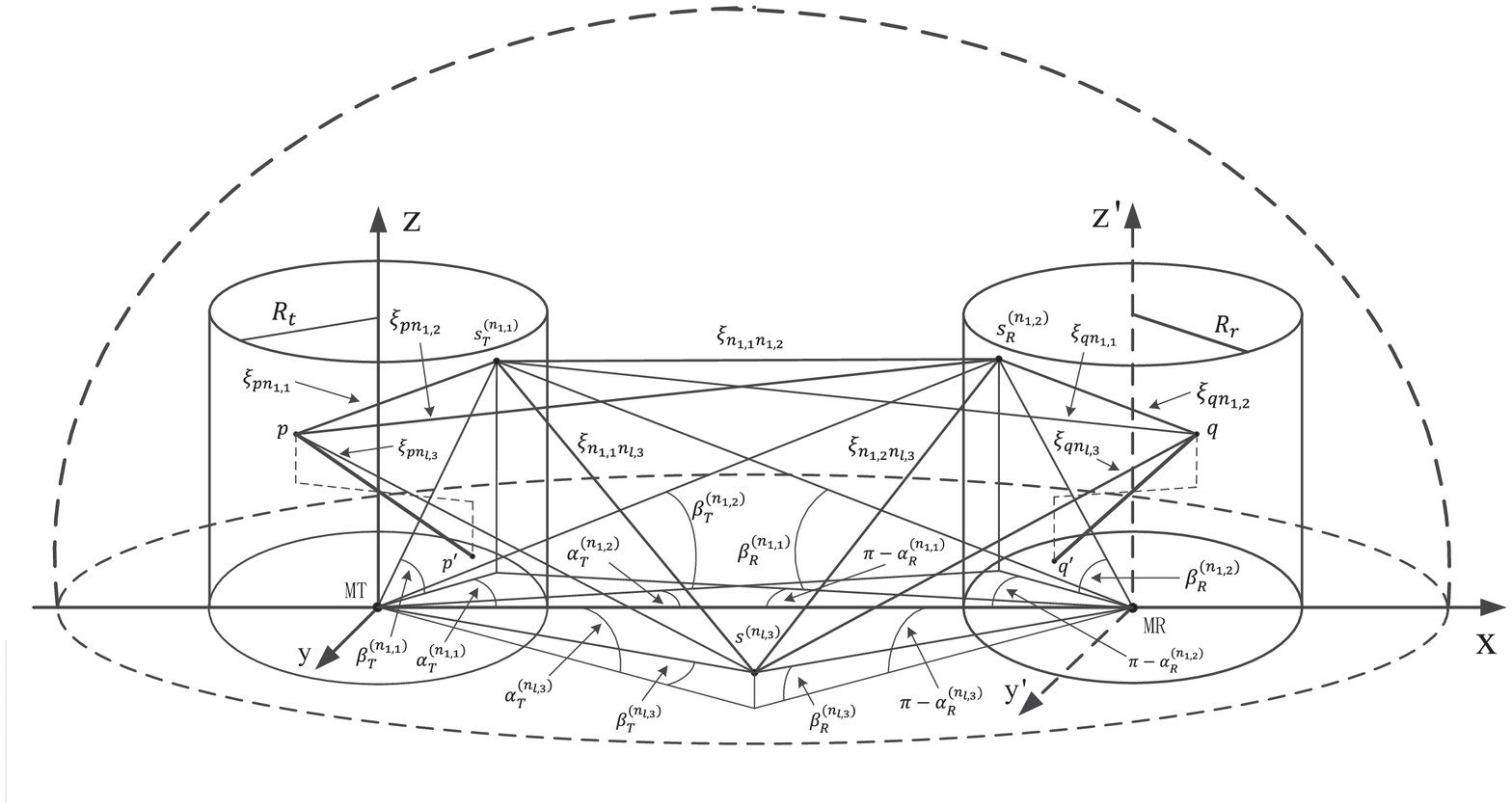,width=9cm} \caption{ Geometric angles and path lengths of the proposed V2V channel model for single- and double-bounced propagation rays. }
\end{center}
\end{figure}

In the proposed channel model, we assume that the MT and MR are located in the same azimuth plane. Thus, the model is mainly applicable for flat road conditions. Similar assumption can be seen in [25] and [27]. However, in reality, the vehicles can be anywhere above, below, or on the actual slope, requiring a more careful analysis to accurately model this V2V propagation condition. For example, the authors of [35] presented path loss channel models for sloped-terrain scenarios, in which the ground reflection was considered in the V2V channels. In this study, the authors introduced four V2V scenarios: (1) two vehicles are located at opposite ends of the slope; (2) one vehicle is on the slope, and the other vehicle is beyond the slope crest; (3) one vehicle is on the slope, and the other is away from the slope at the bottom; (4) both vehicles are on the slope. Figs. 1 and 2 illustrate the geometry of the proposed V2V channel model, which is the combination of line-of-sight (LoS), single-, and double-bounced propagation rays. Here, we use a two-cylinder model to depict moving vehicles (i.e., around the MT or MR). We employ multiple confocal semi-ellipsoid models to mimic stationary roadside environments. In general, we note that most structures in macrocell scenarios (e.g., buildings, highways, urban spaces) have straight vertical surfaces. Thus, we adopt vertical cylinders to model the scattering surfaces represented by moving vehicles [13,19]. Because the heights of the vehicles and pedestrians are similar to those of the transmitter and receiver, we can assume that the scatterers lie on the cylinder model at the MT and MR in the proposed 3D space. To justify this assumption, corresponding comparisons are made between the assumptions of the moving vehicles of the 2D circle and 3D cylinder models. The results show that the power levels of the Doppler spectrum between these models are insignificant. Additionally, we introduce the 3D semi-ellipsoid model because of the following points. (1) For V2V communications, it is acceptable to introduce an ellipse channel model with an MT and MR located at the foci to describe the roadside environments [5]; however, they neglect the transmission signal in the vertical plane. (2) Geometric path lengths between the MT and MR in a 3D semi-ellipsoid V2V channel model continue to change as the transmit azimuth and elevation angles constantly vary. Thus, we can analyze the proposed statistical properties for different path delays as the tap is fixed. (3) We can further analyze the channel statistics for different path delays in different taps. This approach is significantly different from those in previous works [12,27]. To the best of our knowledge, this is the first time that a 3D semi-ellipsoid model is used to mimic V2V channels.

As shown in Figs. 1 and 2, suppose that the MT and MR are equipped with uniform linear array (ULA) $\emph{M}_\emph{T}$ and $\emph{M}_\emph{R}$ omnidirectional antenna elements. The proposed model is also capable of introducing other MIMO geometric antenna systems, such as uniform circular array (UCA), uniform rectangular array (URA), and L-shaped array. The distance between the centers of the MT and MR cylinders are denoted as $\emph{D} = 2f_0$, where $f_0$ designates the half-length of the distance between the two focal points of the ellipse. Let us define $\emph{a}_l$, $\emph{b}_l$, and $\emph{u}_l$ as the semi-major axis of the three dimensions of the $l$th semi-ellipsoid, where $\emph{b}_l = \sqrt{ \emph{a}^2_l - f^2_0 } $. It is assumed that the radius of the cylindrical surface around the MT is denoted as $\emph{R}_{\emph{t}1} \le \emph{R}_{\emph{t}} \le \emph{R}_{\emph{t}2}$. Note that $\emph{R}_{\emph{t}1}$ and $\emph{R}_{\emph{t}2}$ correspond with the respective urban and highway scenarios in [20]. Similarly, at the MR, the radius of the cylindrical surface is denoted as $\emph{R}_{\emph{r}1} \le \emph{R}_{\emph{r}} \le \emph{R}_{\emph{r}2}$. Let Ant$^\emph{T}_\emph{p}$ represent the \emph{p}th ($\emph{p}=1,2,...,\emph{M}_\emph{T}$) antenna of the transmit array, and let Ant$^\emph{R}_\emph{q}$ represent the \emph{q}th ($\emph{q}=1,2,...,\emph{M}_\emph{R}$) antenna of the receive array. The spaces between the two adjacent antenna elements at the MT and MR are denoted as $\delta_\emph{T}$ and $\delta_\emph{R}$, respectively. The orientations of the transmit antenna array in the azimuth plane (relative to the \emph{x}-axis) and elevation plane (relative to the \emph{x-y} plane) are denoted as $\psi_\emph{T}$ and $\theta_\emph{T}$, respectively. Similarly, the orientations at the receiver are denoted as $\psi_\emph{R}$ and $\theta_\emph{R}$, respectively. Here, we assume that there are $\emph{N}_{1\emph{,}1}$ scatterers (moving vehicles) existing on the cylindrical surface around the MT, and the $\emph{n}_{1\emph{,}1}$th ($\emph{n}_{1\emph{,}1} = 1, ... , \emph{N}_{1\emph{,}1}$) scatterer is defined as $\emph{s}^{(\emph{n}_{1\emph{,}1})}_\emph{T}$. $\emph{N}_{1\emph{,}2}$ effective scatterers likewise exist around the MR lying on the cylinder model, and the $\emph{n}_{1\emph{,}2}$th ($\emph{n}_{1\emph{,}2}=1, ... , \emph{N}_{1\emph{,}2}$) scatterer is defined as $\emph{s}^{(\emph{n}_{1\emph{,}2})}_\emph{R}$. For the multiple confocal semi-ellipsoid models, $\emph{N}_{l\emph{,}3}$ scatterers lie on a multiple confocal semi-ellipsoid with the MT and MR located at the foci. The $\emph{n}_{l\emph{,}3}$th ($\emph{n}_{l\emph{,}3}=1, ... , \emph{N}_{l\emph{,}3}$) scatterer is designated as $\emph{s}^{( \emph{n}_{l\emph{,}3} )}$. Although the proposed channel model only takes into account the azimuth and elevation angles in the 3D space, it can also be used in polarized antenna arrays [40], as the polarization angles are considered in the model.

% The proposed 3D non-stationary V2V channel model is operated at 5.4 GHz, with a bandwidth of 50 MHz. Actually, the proposed band is capable of depicting a variety of V2V environments, such as urban and highway scenarios. Moreover, it can also be used to estimate some other V2V band conditions [37]. Note that the proposed band is close to the 5.9 GHz V2V band [21]. However, the difference between 5.4 and 5.9 GHz is 9.3\% ($=$ 0.5/5.4); thus, their propagation channel characteristics do not change significantly. Based on the measurements in [38] and [39], the path loss exponent has a variation of less than 15\% over 1 GHz bandwidth and the delay spread has less than 10\% variation over 8 GHz bandwidth. Here, we could regard these values as an uncertainty of the estimated model parameters at 5.4 GHz, when the goal is to estimate parameter values at 5.9 GHz.

\begin{figure}[t]
\begin{center}
\epsfig{file=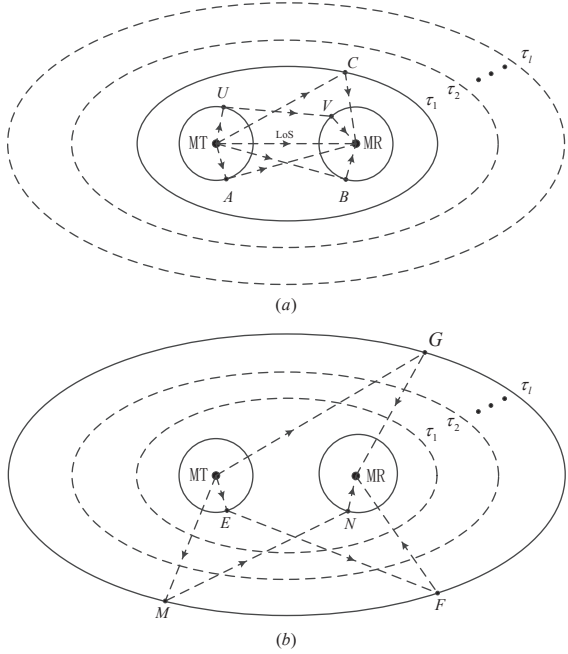,width=8cm} \caption{ The ellipse model describing the path geometry (a) first tap; (b) other taps. }
\end{center}
\end{figure}

In multipath channels, the path length of each wave determines the propagation delay and essentially also the average power of the wave at the MR. In [21], the authors state that the ellipse model forms to a certain extent the physical basis for the modelling of frequency-selective channels. Therefore, when the MT and MR are located in the focus of the ellipse, every wave in the scattering region characterized by the $l$th ellipses undergoes the same discrete propagation delay $\tau_\ell = \tau_0 + \ell \tau$, $\ell=0,1,2,...,\mathcal{L}-1$, where $\tau_0$ denotes the propagation delay of the LoS component, $\tau$ is an infinitesimal propagation delay, and $\mathcal{L}$ is the number of paths with different propagation delays. In particular, the number of paths $\ell$ with different propagation delays exactly corresponds to the number of delay elements required for the tapped-delay-line (TDL) structure of modelling frequency-selective channels. We observe that in real V2V communication scenarios with different contributions of single- and double-bounced rays to the V2V channel statistics, it is necessary to design different taps of the proposed wideband V2V channel model. As mentioned in [36], the tap is strongly related to the delay resolution in V2V channels. Here, let us define $a_l$ as the semi-major of the $l$th ellipse in the azimuth plane. Then, for the next time delay, the semi-major of the $(l+1)$th ellipse in the azimuth plane can be derived as $a_{l+1} = a_l + c\tau / 2$ with $c = 3 \times 10^8$ m/s.

Modelling V2V channels by using a TDL structure with time-variant coefficients gives a deep insight into the channel statistics in the proposed model. In Fig. 3(a), we notice that the received signal for the first tap is composed of an infinite number of delayed and weighted replicas of the transmitted signal in a multipath channel, including direct LoS rays (i.e., MT $\to$ MR), single-bounced rays caused by the scatterers located on either of the two cylinders (i.e., MT $\to \emph{A} \to$ MR and MT $\to \emph{B} \to$ MR) or on the first semi-ellipsoid (i.e., MT $\to \emph{C} \to$ MR), and double-bounced rays generated from the scatterers located on both cylinders (i.e., MT $\to \emph{U} \to \emph{V} \to$ MR). Here, let us define the combination of the above cases as the first tap. Thus, we can analyze the proposed channel characteristics for different time delays, i.e., per-tap channel statistics, which is meaningful for V2V channels. However, for other taps ($l \ge 1$), the link is a superposition of the single-bounced rays that are produced only from the scatterers located on the corresponding semi-ellipsoid (i.e., MT $\to \emph{G} \to$ MR), as well as the double-bounced rays caused by the scatterers from the combined single cylinder (i.e., MT $\to \emph{E} \to \emph{F} \to$ MR) and the corresponding semi-ellipsoid (i.e., MT $\to \emph{M} \to \emph{N} \to$ MR), as shown in Fig. 3(b).

In general, the proposed V2V channel model can be described by matrix $\emph{H}( \emph{t} ) = \big[ \emph{h}_\emph{pq} ( \emph{t}, \tau ) \big]_{ \emph{M}_\emph{T} \times \emph{M}_\emph{R} }$ of size $\emph{M}_\emph{T} \times \emph{M}_\emph{R}$. Therefore, the complex impulse response between the \emph{p}th transmit antenna and \emph{q}th receive antenna in our model can be expressed as $\emph{h}_\emph{pq}( \emph{t}, \tau ) = \sum_{ l=1 }^{ \emph{L}( \emph{t} ) } \omega_l \hspace*{0.04cm} \emph{h}_{ l\emph{,pq} } ( \emph{t} ) \hspace*{0.03cm} \delta ( \tau - \tau_l (\emph{t}) )$, where the subscript $l$ represents the tap number, $\emph{h}_{ l\emph{,pq} } ( \emph{t} ) $ denotes the complex tap coefficient of the Ant$^\emph{T}_\emph{p} \to $ Ant$^\emph{R}_\emph{q}$ link, $\emph{L}( \emph{t} )$ is the total number of taps, $\omega_l$ is the attenuation factors of the $l$th tap, and $\tau_l$ is the corresponding propagation time delays [25].

\subsection{ Proposed 3D channel model description }

Based on the above analysis, the complex tap coefficient for the first tap of the Ant$^\emph{T}_\emph{p} \to $ Ant$^\emph{R}_\emph{q}$ link at the carrier frequency $f_\emph{c}$ can be expressed as [13][27]

\begin{eqnarray}
\emph{h}_{1\emph{,pq}}( \emph{t} ) = \emph{h}^{ \emph{LoS} }_{ 1\emph{,pq} }( \emph{t} ) + \sum_{ i=1 }^{3} \emph{h}^{ \emph{SB}_{1\emph{,}i} }_{ 1\emph{,pq} }( \emph{t} ) + \emph{h}^{ \emph{DB} }_{ 1\emph{,pq} }( \emph{t} )
\end{eqnarray}

\hspace*{-0.475cm} with
\begin{small}
\begin{eqnarray}
\emph{h}^{\emph{LoS}}_{1\emph{,pq}}( \emph{t} ) = \sqrt{ \frac{ \Omega }{ \Omega +1} } e^{ - j 2 \pi f_\emph{c} \xi_\emph{pq} / \emph{c} + j 2\pi \emph{t} \times f_\emph{max}  \cos\big( \alpha^{\emph{LoS}}_\emph{R} - \gamma_\emph{R} \big) \cos\beta^\emph{LoS}_\emph{R}  \hspace*{0.05cm} }
\end{eqnarray}
\end{small}
\begin{eqnarray}
\emph{h}^{\emph{SB}_{1\emph{,}i}}_{1\emph{,pq}} ( \hspace*{-0.425cm} &\emph{t}& \hspace*{-0.425cm} ) = \sqrt{ \frac{ \eta_{\emph{SB}_{1\emph{,}i}} }{ \Omega +1} } \lim_{ \emph{N}_{1\emph{,}i} \to \infty} \sum_{ \emph{n}_{1\emph{,}i} = 1 }^{ \emph{N}_{1\emph{,}i} } \frac{1}{ \sqrt{ \emph{N}_{1\emph{,}i} } } e^{ - j2 \pi f_\emph{c} \xi_{\emph{pq}\emph{,}\emph{n}_{1\emph{,}i} } / \emph{c} } \nonumber \\ [0.2cm]
&\times& \hspace*{-0.25cm} e^{j 2\pi \emph{t} \times f_\emph{max} \cos\big( \alpha^{(\emph{n}_{1\emph{,}i})}_\emph{R} - \gamma_\emph{R} \big) \cos\beta^{(\emph{n}_{1\emph{,}i})}_\emph{R} }, \hspace*{0.15cm} i=1,2,3.
\end{eqnarray}
\begin{eqnarray}
\hspace*{-0.5cm} \emph{h}^{ \emph{DB} }_{1\emph{,pq}}( \emph{t} ) \hspace*{-0.25cm} &=& \hspace*{-0.25cm}  \sqrt{ \frac{ \eta_{\emph{DB} } }{ \Omega +1} } \times \lim_{ \emph{N}_{1\emph{,}1}, \emph{N}_{1\emph{,}2} \to \infty } \sum_{ \emph{n}_{1\emph{,}1}, \emph{n}_{1\emph{,}2} =1}^{ \emph{N}_{1\emph{,}1}, \emph{N}_{1\emph{,}2} } \sqrt{ \frac{1}{ \emph{N}_{1\emph{,}1} \emph{N}_{1\emph{,}2} } } \nonumber \\ [0.2cm]
&\times& \hspace*{-0.25cm} e^{ - j2 \pi f_\emph{c} \xi_{ \emph{pq,}\emph{n}_{1\emph{,}1}\emph{,}\emph{n}_{1\emph{,}2} } / \emph{c} } \nonumber \\ [0.2cm]
&\times& \hspace*{-0.25cm} e^{ j 2\pi \emph{t} \times f_\emph{max} \cos\big( \alpha^{(\emph{n}_{1\emph{,}2})}_\emph{R} - \gamma_\emph{R} \big) \cos\beta^{(\emph{n}_{1\emph{,}2})}_\emph{R} }
\end{eqnarray}

\hspace*{-0.475cm} where $\xi_{\emph{pq}\emph{,}\emph{n}_{1\emph{,}i} } = \xi_{\emph{p}\emph{n}_{1\emph{,}i}} + \xi_{\emph{q}\emph{n}_{1\emph{,}i}} $ and $\xi_{ \emph{pq,}\emph{n}_{1\emph{,}1}\emph{,}\emph{n}_{1\emph{,}2} } = \xi_{ \emph{p}\emph{n}_{1\emph{,}1} } + \xi_{ \emph{n}_{1\emph{,}1} \emph{n}_{1\emph{,}2} } + \xi_{ \emph{q}\emph{n}_{1\emph{,}2} }$ denote the travel distance of the waves through the link Ant$^\emph{T}_\emph{p} \to \emph{s}^{(\emph{n}_{1\emph{,}i})} \to$ Ant$^\emph{R}_\emph{q}$ and Ant$^\emph{T}_\emph{p} \to \emph{s}^{(\emph{n}_{1\emph{,}1})}_\emph{T} \to \emph{s}^{(\emph{n}_{1\emph{,}2})}_\emph{R} \to$ Ant$^\emph{R}_\emph{q}$, respectively. Here, $\Omega$ denotes the Rice factor and $f_\emph{max}$ is the maximum Doppler frequency with respect to the MR [11]. $\alpha^\emph{LoS}_\emph{R}$ and $\beta^\emph{LoS}_\emph{R}$ denote the AAoA and EAoA of the LoS path, respectively. For the NLoS rays, the symbol $\alpha^{ (\emph{n}_{1\emph{,}1}) }_\emph{R}$ represents the AAoA of the wave scattered from the effective scatterer $\emph{s}^{(\emph{n}_{1\emph{,}1})}_\emph{T}$ around the MT, whereas $\alpha^{ (\emph{n}_{1\emph{,}2}) }_\emph{R}$ represents the AAoA of the wave scattered from the scatterer $\emph{s}^{(\emph{n}_{1\emph{,}2})}_\emph{R}$ around the MR. Similarly, $\beta^{ (\emph{n}_{1\emph{,}1}) }_\emph{R}$ and $\beta^{ (\emph{n}_{1\emph{,}2}) }_\emph{R}$ denote the EAoAs of the waves scattered from the scatterer $\emph{s}^{(\emph{n}_{1\emph{,}1})}_\emph{T}$ and $\emph{s}^{(\emph{n}_{1\emph{,}2})}_\emph{R}$, respectively. On the other hand, $\alpha^{ (\emph{n}_{1\emph{,}3}) }_\emph{R}$ and $\beta^{ (\emph{n}_{1\emph{,}3}) }_\emph{R}$ denote the AAoA and EAoA of the waves scattered from the scatterer $\emph{s}^{(\emph{n}_{1\emph{,}3})}$ in the semi-ellipsoid model for the first tap. It is evident that the MT and MR are both in motion, we herein assume that the MR moves in a relative direction to the MT with the principles of relative motion.\footnotemark[2] Similar work can be seen in [2] and [5]. In this case, different channel characteristics can be described by adjusting the related model parameters. Here, we assume that the MR moves in an arbitrary direction, $\gamma_\emph{R}$, with a constant velocity of $\emph{v}_\emph{R}$ at time instant $\emph{t}$ in the azimuth plane. Furthermore, energy-related parameters $\eta_{ \emph{SB}_{1\emph{,}i} }$ and $\eta_{ \emph{DB} }$ specify the numbers of the single- and double-bounced rays respectively contribute to the total scattered power, which can be normalized to satisfy $\sum_{i=1}^{3} \eta_{ \emph{SB}_{1\emph{,}i} } + \eta_\emph{DB} = 1$ for brevity [11,21]. However, as shown in Fig. 2, for other taps ($l \ge$ 1), the complex tap coefficient of the Ant$^\emph{T}_\emph{p} \to $ Ant$^\emph{R}_\emph{q}$ link can be derived as
\begin{eqnarray}
\emph{h}_{l\emph{,pq}}( \emph{t} ) = \emph{h}^{ \emph{SB}_{l\emph{,}3} }_{l\emph{,pq}} ( \emph{t} ) + \emph{h}^{ \emph{DB}_{l\emph{,}1} }_{l\emph{,pq}} ( \emph{t} ) + \emph{h}^{ \emph{DB}_{ l\emph{,}2 } }_{ l\emph{,pq} } ( \emph{t} )
\end{eqnarray}

\footnotetext[2]{Although the existing channel models, where the MT and MR are both in motion with same or opposite directions, seem more reasonable to reflect the actual vehicular environments. In reality, these models cannot study the effects of arbitrary moving directions on the channel statistics, which are meaningful for V2V channel.}

\hspace*{-0.475cm} with
\begin{eqnarray}
\emph{h}^{\emph{SB}_{l\emph{,}3 } }_{l\emph{,pq}} ( \emph{t} ) \hspace*{-0.25cm} &=& \hspace*{-0.25cm} \sqrt{ \eta_{\emph{SB}_{l\emph{,}3 } } } \lim_{ \emph{N}_{l\emph{,}3 } \to \infty} \sum_{\emph{n}_{l\emph{,}3}=1}^{ \emph{N}_{l\emph{,}3 } } \frac{1}{ \sqrt{ \emph{N}_{l\emph{,}3}} } e^{ - j 2 \pi f_\emph{c} \xi_{ \emph{pq}\emph{,n}_{l\emph{,}3 } } / \emph{c} } \nonumber \\ [0.2cm]
&\times& \hspace*{-0.25cm} e^{ j 2\pi \emph{t} \times f_\emph{max} \cos\big( \alpha^{( \emph{n}_{l\emph{,}3 } )}_\emph{R} - \gamma_\emph{R} \big) \cos\beta^{( \emph{n}_{l\emph{,}3 } )}_\emph{R} \hspace*{0.05cm} }
\end{eqnarray}
\begin{eqnarray}
\hspace*{-0.5cm} \emph{h}^{ \emph{DB}_{l\emph{,}1} }_{ l\emph{,pq} } ( \emph{t} ) \hspace*{-0.25cm} &=& \hspace*{-0.25cm} \sqrt{ \eta_{\emph{DB}_{ l\emph{,}1 } } } \lim_{ \emph{N}_{1\emph{,}1}, \emph{N}_{l\emph{,}3} \to \infty } \sum_{ \emph{n}_{1\emph{,}1}, \emph{n}_{l\emph{,}3} =1}^{ \emph{N}_{1\emph{,}1}, \emph{N}_{l\emph{,}3} } \sqrt{ \frac{1}{ \emph{N}_{1\emph{,}1} \emph{N}_{l\emph{,}3} } } \nonumber \\ [0.2cm]
&\times& \hspace*{-0.25cm} e^{ - j2 \pi f_\emph{c} \xi_{ \emph{pq,}\emph{n}_{1\emph{,}1}\emph{,}\emph{n}_{l\emph{,}3} } / \emph{c} } \nonumber \\ [0.2cm]
&\times& \hspace*{-0.25cm} e^{ j 2\pi \emph{t} \times f_\emph{max} \cos\big( \alpha^{(\emph{n}_{l\emph{,}3})}_\emph{R} - \gamma_\emph{R} \big) \cos\beta^{(\emph{n}_{l\emph{,}3})}_\emph{R} \hspace*{0.05cm} }
\end{eqnarray}
\begin{eqnarray}
\hspace*{-0.5cm} \emph{h}^{ \emph{DB}_{l\emph{,}2} }_{ l\emph{,pq} } ( \emph{t} ) \hspace*{-0.25cm} &=& \hspace*{-0.25cm} \sqrt{ \eta_{\emph{DB}_{l\emph{,}2 } } } \times  \lim_{ \emph{N}_{l\emph{,}3}, \emph{N}_{1\emph{,}2} \to \infty } \sum_{ \emph{n}_{l\emph{,}3}, \emph{n}_{1\emph{,}2} =1}^{ \emph{N}_{l\emph{,}3}, \emph{N}_{1\emph{,}2} } \sqrt{ \frac{1}{ \emph{N}_{l\emph{,}3} \emph{N}_{1\emph{,}2} } } \nonumber \\ [0.2cm]
&\times& \hspace*{-0.25cm}e^{ - j2 \pi f_\emph{c} \xi_{ \emph{pq,}\emph{n}_{l\emph{,}3}\emph{,}\emph{n}_{1\emph{,}2} } / \emph{c} } \nonumber \\ [0.2cm]
&\times& \hspace*{-0.25cm} e^{ j 2\pi \emph{t} \times f_\emph{max} \cos\big( \alpha^{(\emph{n}_{1\emph{,}2})}_\emph{R} - \gamma_\emph{R} \big)  \cos\beta^{(\emph{n}_{1\emph{,}2})}_\emph{R}  }
\end{eqnarray}

\hspace*{-0.475cm} where $\xi_{ \emph{pq}\emph{,n}_{l\emph{,}3 } } = \xi_{ \emph{p}\emph{n}_{l\emph{,}3 } } + \xi_{ \emph{q}\emph{n}_{l\emph{,}3 } }$, $\xi_{ \emph{pq,}\emph{n}_{1\emph{,}1}\emph{,}\emph{n}_{l\emph{,}3 } } = \xi_{ \emph{p}\emph{n}_{1\emph{,}1} } + \xi_{ \emph{n}_{1\emph{,}1} \emph{n}_{l\emph{,}3 } } + \xi_{ \emph{q}\emph{n}_{l\emph{,}3 } } $, and $\xi_{ \emph{pq,}\emph{n}_{l\emph{,}3}\emph{,}\emph{n}_{1\emph{,}2} } = \xi_{ \emph{p}\emph{n}_{l\emph{,}3 } } + \xi_{ \emph{n}_{l\emph{,}3 } \emph{n}_{1\emph{,}2} } + \xi_{ \emph{q}\emph{n}_{1\emph{,}2 } }$ denote the travel distance of the waves through the link Ant$^\emph{T}_\emph{p} \to \emph{s}^{(\emph{n}_{l\emph{,}3 })} \to$ Ant$^\emph{R}_\emph{q}$, Ant$^\emph{T}_\emph{p} \to \emph{s}^{(\emph{n}_{1\emph{,}1})}_\emph{T} \to \emph{s}^{(\emph{n}_{l\emph{,}3})} \to$ Ant$^\emph{R}_\emph{q}$, and Ant$^\emph{T}_\emph{p} \to \emph{s}^{(\emph{n}_{l\emph{,}3})} \to \emph{s}^{(\emph{n}_{1\emph{,}2})}_\emph{R} \to$ Ant$^\emph{R}_\emph{q}$, respectively. $\alpha^{ (\emph{n}_{l\emph{,}3}) }_\emph{R}$ and $\beta^{ (\emph{n}_{l\emph{,}3}) }_\emph{R}$ denote the AAoA and EAoA of the waves scattered from the scatterer $\emph{s}^{(\emph{n}_{l\emph{,}3})}$ in the $l$th semi-ellipsoid model for other taps. Similar to the above case, energy-related parameters $\eta_{\emph{SB}_{l\emph{,}3}}$ and $\eta_{\emph{DB}_{l\emph{,}1} }$ ($\eta_{\emph{DB}_{l\emph{,}2} }$) specify the amount that the single- and double-bounced rays respectively contribute to the total scattered power, which can be normalized to satisfy $\eta_{ \emph{SB}_{l\emph{,}3} } + \eta_{\emph{DB}_{l\emph{,}1} } + \eta_{\emph{DB}_{l\emph{,}2} } = 1$ for brevity. In addition, because the derivations of the condition that guarantees the fulfillment of the TDL structure are the same, we only detail the derivation of the condition for the second tap.

As introduced in [12] and [27], we note that the impulse response of the proposed model is related to the scattered power in V2V channels. Therefore, it is important to define the received scattered power in different taps and different V2V scenarios (i.e., highway and urban scenarios) in the proposed non-stationary channel model. In short, for the first tap, the single-bounced rays are caused by the scatterers located on either of the two cylinders or the first semi-ellipsoid, while the double-bounced rays are generated from the scatterers located on the both cylinders, as shown in Fig. 2. For highway scenarios (i.e., $\emph{R}_\emph{t} = \emph{R}_{\emph{t}2}$ and $\emph{R}_\emph{r} = \emph{R}_{\emph{r}2}$), the higher relative movement of the vehicles results in a higher Doppler frequency; moreover, the value of $\Omega$ is always large because the LoS component can bear a significant amount of power. Additionally, the received scattered power is mainly from waves reflected by the stationary roadside environments described by the scatterers located on the first semi-ellipsoid. The moving vehicles represented by the scatterers located on the two cylinders are more likely to be single-bounced, rather than double-bounced. This indicates that $\eta_{\emph{SB}_{1\emph{,}3}} > \max \{ \eta_{\emph{SB}_{1\emph{,}1}}, \eta_{\emph{SB}_{1\emph{,}2}} \} > \eta_\emph{DB} $. For urban scenarios (i.e., $\emph{R}_\emph{t} = \emph{R}_{\emph{t}1}$ and $\emph{R}_\emph{r} = \emph{R}_{\emph{r}1}$), the lower relative movement of the vehicles results in a lower Doppler frequency; moreover, the value of $\Omega$ is smaller than that in the highway scenarios. Additionally, the double-bounced rays of the two-cylinder model can bear more energy than the single-bounced rays of the two-cylinder and semi-ellipsoid models, i.e., $\eta_\emph{DB} > \max \{ \eta_{\emph{SB}_{1\emph{,}1}}, \eta_{\emph{SB}_{1\emph{,}2}}, \eta_{\emph{SB}_{1\emph{,}3}} \}$.

However, for the second tap, it is assumed that the single-bounced rays are produced only from the static scatterers located on the corresponding semi-ellipsoid, while the double-bounced rays are caused by the scatterers from the combined one cylinder (either of the two cylinders) and the corresponding semi-ellipsoid [21,26,27]. Note that, in the proposed TDL structure, the double-bounced rays in the first tap must be smaller in distance than the single-bounced rays on the second semi-ellipsoid, i.e., $\max \{\emph{R}_\emph{t}, \emph{R}_\emph{r}\} < \min \{\emph{a}_2 - \emph{a}_1\}$. It is stated in [41] that the delay resolution is approximately the inverse of bandwidth and therefore, we assume that the delay resolution in the proposed model is 20 ns for 50 MHz. In this paper, we define different time delays with the different ellipses. Thus, the second ellipse should produce at least 6 m excess path length than the first ellipse, i.e., $2\emph{a}_2 - 2\emph{a}_1 = \emph{c} \tau$ with $\tau = 20$ ns. In this case, the proposed channel statistics for different time delays, i.e., per-tap statistics, can be investigated. For highway scenarios, the received scattered power is mainly from waves reflected by the stationary roadside environments described by the scatterers located on the semi-ellipsoid, i.e., $\eta_{\emph{SB}_{l\emph{,}3}} > \max \{ \eta_{\emph{DB}_{l\emph{,}1}}, \eta_{\emph{DB}_{l\emph{,}2}} \}$. For urban scenarios, the double-bounced rays from the combined single cylinder and semi-ellipsoid models can bear more energy than the single-bounced rays of the semi-ellipsoid model, i.e., $\min \{ \eta_{\emph{DB}_{l\emph{,}1}}, \eta_{\emph{DB}_{l\emph{,}2}} \} > \eta_{\emph{SB}_{l\emph{,}3}}$.

\begin{figure}[t]
\begin{center}
\epsfig{file=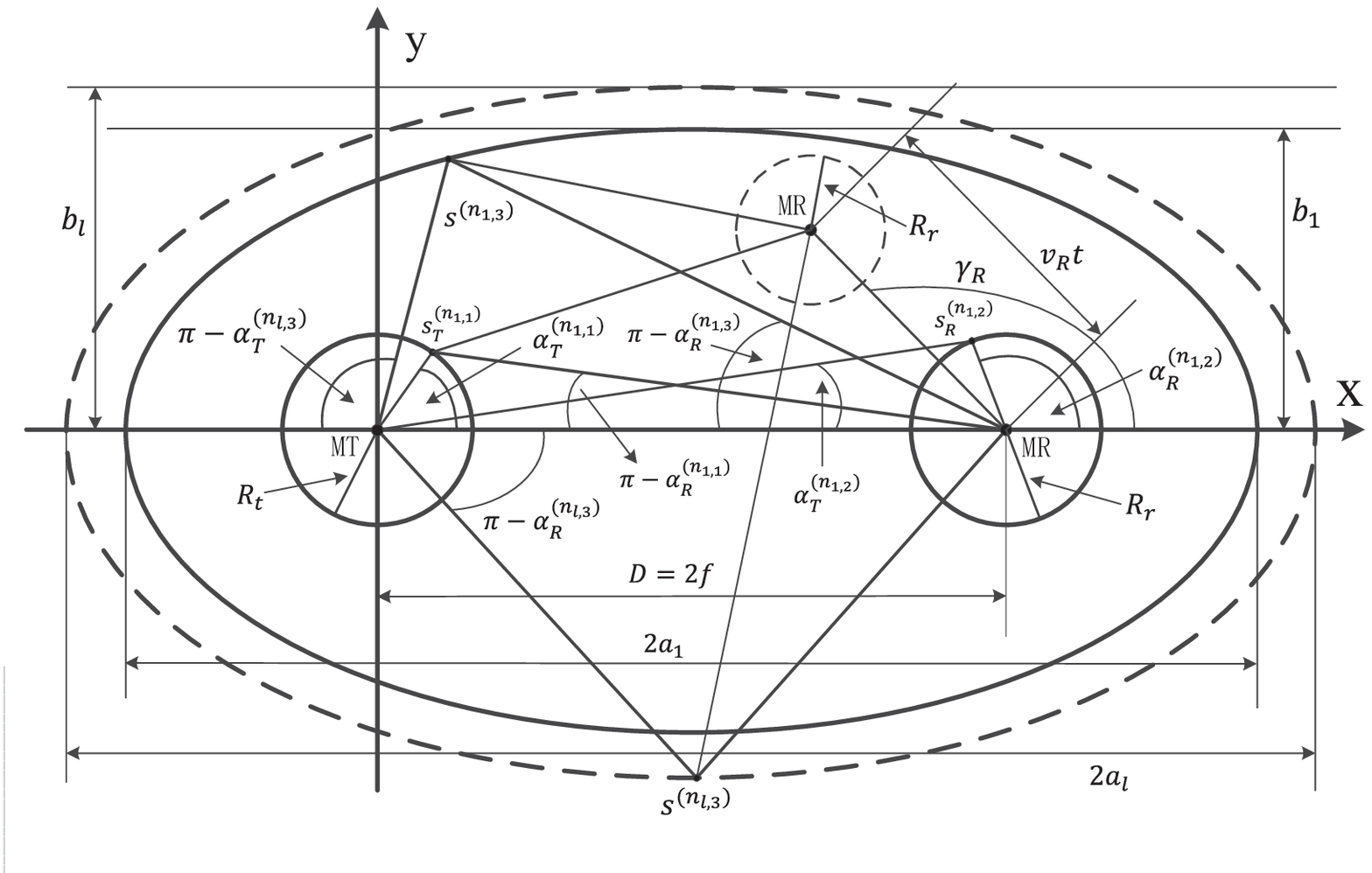,width=9.15cm} \caption{ Top view of the geometric angles and path lengths in the proposed non-stationary V2V channel model. }
\end{center}
\end{figure}

\subsection{ Non-stationary time-variant parameters }

To describe the non-stationarity of the proposed 3D wideband channel model, we introduce a V2V communication scenario, as illustrated in Fig. 4. The figure shows the geometric properties and moving statistics of the proposed model in the azimuth plane. In this case, owing to overly complex issues, the corresponding 3D figure with MIMO antennas is omitted for brevity. For V2V scenarios, the geometric paths lengths will be time-variant because of the relative movement between the MT and MR. Consequently, $\xi_\emph{pq}$, $\xi_{\emph{p}\emph{n}_{1\emph{,}2}}$, $\xi_{\emph{q}\emph{n}_{1\emph{,}1}}$, $\xi_{\emph{q}\emph{n}_{l\emph{,}3 }}$, and $\xi_{\emph{n}_{1\emph{,}1}\emph{n}_{1\emph{,}2}}$ can be replaced by $\xi_\emph{pq}( \emph{t} )$, $\xi_{\emph{p}\emph{n}_{1\emph{,}2}}( \emph{t} )$, $\xi_{\emph{q}\emph{n}_{1\emph{,}1}}( \emph{t} )$, $\xi_{\emph{q}\emph{n}_{l\emph{,}3 }}( \emph{t} )$, and $\xi_{\emph{n}_{1\emph{,}1}\emph{n}_{1\emph{,}2}} ( \emph{t} )$, respectively. However, in Fig. 4, note that the distances $\xi_{\emph{p}\emph{n}_{1\emph{,}1}}$, $\xi_{\emph{p}\emph{n}_{l\emph{,}3 }}$, and $\xi_{\emph{q}\emph{n}_{1\emph{,}2}}$ have no related to the non-stationary properties, i.e., $\xi_{\emph{p}\emph{n}_{1\emph{,}1}}( \emph{t} ) = \xi_{\emph{p}\emph{n}_{1\emph{,}1}}$, $\xi_{\emph{p}\emph{n}_{l\emph{,}3 }}( \emph{t} ) = \xi_{\emph{p}\emph{n}_{l\emph{,}3 }} $, and $\xi_{\emph{q}\emph{n}_{1\emph{,}2}} ( \emph{t} ) = \xi_{\emph{q}\emph{n}_{1\emph{,}2}} $. In general, it is clearly observed that the MR is relatively far from the MT in the proposed V2V communication environments. Thus, we can make the following assumptions: $\min \{\emph{R}_\emph{t}, \emph{R}_\emph{r}, \emph{u}-f \} \gg \max \{\delta_\emph{T}, \delta_\emph{R} \}$, $\emph{D} \gg \max \{\delta_\emph{T}, \delta_\emph{R} \}$, and the approximation $\sqrt{\emph{x}+1} \approx 1 + \emph{x}/2$ is used for small \emph{x}. Accordingly, based on the law of cosines in appropriate triangles and small angle approximations (i.e., $\sin \emph{x} \approx \emph{x}$ and $\cos \emph{x} \approx 1$ for small \emph{x}) [12,27], the corresponding time-variant geometric path lengths at relative moving time instant \emph{t} can be approximated as

\begin{small}
\begin{eqnarray}
\xi_\emph{pq} ( \emph{t} ) \approx \sqrt{ \big( \emph{D} - \delta_\emph{Tx} \big)^2 + \big( \emph{v}_\emph{R} \emph{t} \big)^2 - 2\big( \emph{D} - \delta_\emph{Tx} \big) \emph{v}_\emph{R} \emph{t} \cos\big( \alpha^\emph{LoS}_\emph{R} - \gamma_\emph{R} \big) }
\end{eqnarray}
\end{small}
\begin{small}
\begin{eqnarray}
\xi_{\emph{p}\emph{n}_{1\emph{,}1}} ( \emph{t} ) \approx \emph{R}_\emph{t} \hspace*{-0.25cm} &-& \hspace*{-0.25cm} \Big[ \delta_\emph{Tx} \cos\alpha^{(\emph{n}_{1\emph{,}1})}_\emph{T} \cos\beta^{(\emph{n}_{1\emph{,}1})}_\emph{T} \nonumber \\[0.15cm]
\hspace*{-0.25cm} &+& \hspace*{-0.25cm} \delta_\emph{Ty} \sin\alpha^{(\emph{n}_{1\emph{,}1})}_\emph{T} \cos\beta^{(\emph{n}_{1\emph{,}1})}_\emph{T} + \delta_\emph{Tz} \sin\beta^{(\emph{n}_{1\emph{,}1})}_\emph{T} \Big]
\end{eqnarray}
\end{small}
\begin{small}
\begin{eqnarray}
\xi_{\emph{p}\emph{n}_{1\emph{,}2}} ( \emph{t} ) \approx \bigg[ \Big( \emph{D} \hspace*{-0.25cm} &-& \hspace*{-0.25cm} \emph{Q}_\emph{p} \cos\theta_\emph{T} \Big)^2 + \big( \emph{v}_\emph{R} \emph{t} \big)^2 \nonumber \\[0.15cm]
\hspace*{-0.25cm} &-& \hspace*{-0.25cm} 2 \Big( \emph{D} - \emph{Q}_\emph{p} \cos\theta_\emph{T} \Big) \emph{v}_\emph{R} \emph{t} \cos\big( \alpha^{(\emph{n}_{1\emph{,}2})}_\emph{R} - \gamma_\emph{R} \big) \bigg]
\end{eqnarray}
\end{small}
\begin{small}
\begin{eqnarray}
\xi_{\emph{p}\emph{n}_{l\emph{,}3}} ( \emph{t} ) \approx \frac{ 2 \emph{a}^2_l \emph{b}^2_l \emph{u}^2_l }{ \xi_{l,3} }  - \delta_\emph{T} \cos\theta_\emph{T} \Big[ \emph{R}_\emph{r} / \emph{D} \sin\psi_\emph{T} \sin\alpha^{(\emph{n}_{l\emph{,}3})}_\emph{T} + \cos\psi_\emph{T}  \Big]
\end{eqnarray}
\end{small}
\begin{small}
\begin{eqnarray}
\xi_{\emph{q}\emph{n}_{1\emph{,}2}} ( \emph{t} ) \approx \emph{R}_\emph{r} \hspace*{-0.25cm} &-& \hspace*{-0.25cm}   \Big[  \delta_\emph{Rx} \cos\alpha^{(\emph{n}_{1\emph{,}2})}_\emph{R} \cos\beta^{(\emph{n}_{1\emph{,}2})}_\emph{R}  \nonumber \\[0.15cm]
\hspace*{-0.25cm} &+& \hspace*{-0.25cm} \delta_\emph{Ry} \sin\alpha^{(\emph{n}_{1\emph{,}2})}_\emph{R} \cos\beta^{(\emph{n}_{1\emph{,}2})}_\emph{R} + \delta_\emph{Rz} \sin\beta^{(\emph{n}_{1\emph{,}2})}_\emph{R} \Big]
\end{eqnarray}
\end{small}
\begin{small}
\begin{eqnarray}
\xi_{\emph{q}\emph{n}_{1\emph{,}1}} ( \emph{t} ) \approx \bigg[ \Big( \emph{D} \hspace*{-0.25cm} &-& \hspace*{-0.25cm} \emph{Q}_\emph{q} \cos\theta_\emph{R} \Big)^2 + \big( \emph{v}_\emph{R} \emph{t} \big)^2 \nonumber \\[0.15cm]
\hspace*{-0.25cm} &-& \hspace*{-0.25cm} 2 \Big( \emph{D} - \emph{Q}_\emph{q} \cos\theta_\emph{R} \Big) \emph{v}_\emph{R} \emph{t} \cos\big( \alpha^{(\emph{n}_{1\emph{,}1})}_\emph{R} - \gamma_\emph{R} \big) \bigg]
\end{eqnarray}
\end{small}
\begin{small}
\begin{eqnarray}
\xi_{\emph{q}\emph{n}_{l\emph{,}3}} ( \emph{t} ) \approx  \bigg[ \xi^2_{\emph{p}\emph{n}_{l\emph{,}3}} ( \emph{t} ) \hspace*{-0.25cm} &\sin& \hspace*{-0.35cm}^2\beta^{(\emph{n}_{l\emph{,}3})}_\emph{T}  + \xi^2_\emph{R} \nonumber \\[0.15cm]
\hspace*{-0.25cm} &+& \hspace*{-0.25cm} \big( \emph{v}_\emph{R} \emph{t} \big)^2 + 2 \emph{D} \xi_\emph{R} \cos\big( \gamma_\emph{R} + \alpha^{(\emph{n}_{l\emph{,}3})}_\emph{R} \big) \bigg]
\end{eqnarray}
\end{small}
\begin{small}
\begin{eqnarray}
\xi_{ \emph{n}_{1\emph{,}1}\emph{n}_{1\emph{,}2} } ( \emph{t} ) \approx \sqrt{ \emph{D}^2 + \big( \emph{v}_\emph{R} \emph{t} \big)^2 - 2 \emph{D} \emph{v}_\emph{R} \emph{t} \cos \big( \alpha^\emph{LoS}_\emph{R} - \gamma_\emph{R} \big) }
\end{eqnarray}
\end{small}

\hspace*{-0.6cm} where $\delta_\emph{Tx} = \delta_\emph{T} \cos\theta_\emph{T} \cos\psi_\emph{T}$, $\delta_\emph{Ty} = \delta_\emph{T} \cos\theta_\emph{T} \sin\psi_\emph{T}$, $\delta_\emph{Tz} = \delta_\emph{T} \sin\theta_\emph{T} $, $\delta_\emph{Rx} = \delta_\emph{R} \cos\theta_\emph{R} \cos\psi_\emph{R}$, $\delta_\emph{Ry} = \delta_\emph{R} \cos\theta_\emph{R} \sin\psi_\emph{R}$, $\delta_\emph{Rz} = \delta_\emph{R} \sin\theta_\emph{R} $, $\emph{Q}_\emph{p} = \delta_\emph{T} \cos\theta_\emph{T} \big[ \emph{R}_\emph{r} / \emph{D} \sin\psi_\emph{T} \sin\alpha^{(\emph{n}_{1\emph{,}2})}_\emph{R} + \cos\psi_\emph{T} \big]$, $\xi_{l,3} = \emph{b}^2_l \emph{u}^2_l \cos^2 \beta^{(\emph{n}_{l\emph{,}3})}_\emph{T}  \cos^2 \alpha^{(\emph{n}_{l\emph{,}3})}_\emph{T}  +  \emph{a}^2_l \emph{u}^2_l \cos^2 \beta^{(\emph{n}_{l\emph{,}3})}_\emph{T}  \sin^2 \alpha^{(\emph{n}_{l\emph{,}3})}_\emph{T}  + \emph{a}^2_l \emph{b}^2_l \sin^2 \beta^{(\emph{n}_{l\emph{,}3})}_\emph{T}$, $\xi_\emph{R} = \big[ \emph{D}^2 + \xi^2_{\emph{p}\emph{n}_{l\emph{,}3}} ( \emph{t} ) \cos^2\alpha^{(\emph{n}_{l\emph{,}3})}_\emph{T} - 2 \emph{D} \xi_{\emph{p}\emph{n}_{l\emph{,}3}} ( \emph{t} ) \cos\beta^{(\emph{n}_{l\emph{,}3})}_\emph{T} \cos\alpha^{(\emph{n}_{l\emph{,}3})}_\emph{T} \big]$, and $\emph{Q}_\emph{q} = \delta_\emph{R} \cos\theta_\emph{R} \big[ \emph{R}_\emph{t} / \emph{D} \sin\psi_\emph{R} \sin\alpha^{(\emph{n}_{1\emph{,}1})}_\emph{T} - \cos\psi_\emph{R} \big]$.

To jointly consider the impact of the azimuth and elevation angles on channel statistics, several scatterer distributions, such as uniform, Gaussian, Laplacian, and von Mises, were used in prior work. Here, we adopt the von Mises probability density function (PDF) to characterize the distribution of scatterers in the proposed V2V channel. Thus, the von Mises PDF is derived as

\begin{eqnarray}
p (\alpha^{( \emph{n}_{l\emph{,}i} )}_\emph{R}, \beta^{( \emph{n}_{l\emph{,}i} )}_\emph{R})\hspace*{-0.25cm}& = & \hspace*{-0.25cm} \frac{ \emph{k} \cos^{( \emph{n}_{l\emph{,}i} )}_\emph{R} }{ 4\pi\sinh \emph{k} } \times e^{ \emph{k} \cos\beta_0 \cos\beta^{( \emph{n}_{l\emph{,}i} )}_\emph{R} \cos\big(\alpha^{( \emph{n}_{l\emph{,}i} )}_\emph{R} -\alpha_0 \big) } \nonumber \\[0.15cm]
\hspace*{-0.25cm} &\times& \hspace*{-0.25cm} e^{ \emph{k} \sin\beta_0 \sin\beta^{( \emph{n}_{l\emph{,}i} )}_\emph{R} }
\end{eqnarray}

\hspace*{-0.475cm} with $\alpha^{( \emph{n}_{l\emph{,}i} )}_\emph{R}$ and $\beta^{( \emph{n}_{l\emph{,}i} )}_\emph{R} \in [-\pi, \pi)$, $\alpha_0 \in [-\pi, \pi)$. In addition, $\beta_0 \in [-\pi, \pi)$ denotes the mean values of the azimuth angle $\alpha^{( \emph{n}_{l\emph{,}i} )}_\emph{R}$ and elevation angle $\beta^{( \emph{n}_{l\emph{,}i} )}_\emph{R}$ at the receiver, respectively. In addition, \emph{k} (\emph{k} $\ge$ 0) is a real-valued parameter that controls the angles spread of $\alpha_0$ and $\beta_0$ [12].

As previously mentioned, different channel characteristics can be described by adjusting the proposed model parameters. For example, it is apparent that when we do not take the roadside environments into account, the proposed model tends to the Zajic model [13,19]. However, the proposed model is suitable for the previous 3D stationary semi-ellipsoid channels as $\emph{t} = 0$, as shown in [31] and [32]. In this case, our channel can be degenerated into a 2D elliptical channel as model parameter $\emph{u}_l$ is equal to zero. On the other hand, when we set $\emph{t} \neq 0$, our model can can also be used to depict non-stationary V2V channels, such as the Ghazal model [26] and Yuan model [27]. Likewise, the proposed model describes other models in previous work; we omit them for brevity.

\section{ Proposed Theoretical Model Statistical Properties }

\subsection{ Space CFs }

In general, CF is an important statistic in designing communication link that characterizes how fast a wireless channel changes with respect to time, movement, or frequency [25]. The specific CF that is of interest in this paper is the space CF, which measures the spatial statistics of the proposed V2V channel. It is stated in [27] that the spatial correlation properties of two arbitrary channel impulse responses $\emph{h}_\emph{pq} ( \emph{t} )$ and $\emph{h}_\emph{p'q'} ( \emph{t} )$ of a MIMO V2V channel are completely determined by the correlation properties of $\emph{h}_{ l,\emph{pq} } ( \emph{t} )$ and $\emph{h}_{ l,\emph{p'q'} } ( \emph{t} )$ in each tap, so that no correlations exist between the underlying processes in different taps. Therefore, the normalized time-variant space CF can be expressed as [36]

\begin{eqnarray}
\rho_{ \emph{h}_{ l\emph{,pq} }, \emph{h}_{ l\emph{,p'q'} } } \big( \emph{t}, \tau \big) = \frac{ \emph{E} \Big[ \emph{h}_{ l\emph{,pq} } \big( \emph{t} \big) \emph{h}^*_{ l\emph{,p'q'} } \big( \emph{t} + \tau \big) \Big] }{ \sqrt{ \emph{E} \Big[ \big| \emph{h}_{ l\emph{,pq} } ( \emph{t} ) \big|^2 \Big] \emph{E} \Big[ \big| \emph{h}_{ l\emph{,p'q'} } ( \emph{t} + \tau ) \big|^2 \Big] } }
\end{eqnarray}

\hspace*{-0.55cm} where ($\cdot$)$^*$ denotes the complex conjugate operation and $\emph{E} [\cdot]$ is the expectation operation. Because the LoS, single-, and double-bounced rays are independent of each other, the channel response for the first tap can be expressed as
\begin{eqnarray}
\rho_{ \emph{h}_{ 1\emph{,pq} }, \emph{h}_{ 1\emph{,p'q'} } } \big( \emph{t}, \tau \big) \hspace*{-0.25cm}& = & \hspace*{-0.25cm} \rho^{ \emph{LoS} }_{ \emph{h}_{1\emph{,pq}},\emph{h}_{1\emph{,p'q'}} } \big( \emph{t}, \tau \big) + \sum_{i=1}^{3} \rho^{ \emph{SB}_{1\emph{,}i} }_{ \emph{h}_{1\emph{,pq}},\emph{h}_{1\emph{,p'q'}} } \big( \emph{t}, \tau \big) \nonumber \\ [0.25cm]
\hspace*{-0.25cm}& + & \hspace*{-0.25cm} \rho^\emph{DB}_{ \emph{h}_{1\emph{,pq}},\emph{h}_{1\emph{,p'q'}} } \big( \emph{t}, \tau \big)
\end{eqnarray}

However, for other taps, according to (18), we have the time-variant space CF as

\begin{eqnarray}
\rho_{ \emph{h}_{ l\emph{,pq} }, \emph{h}_{ l\emph{,p'q'} } } \big( \emph{t}, \tau \big) \hspace*{-0.25cm}& = & \hspace*{-0.25cm} \rho^{ \emph{SB}_{l\emph{,}3} }_{ \emph{h}_{ l\emph{,pq}}, \emph{h}_{ l\emph{,p'q'}} } \big( \emph{t}, \tau \big) + \rho^{ \emph{DB}_{l\emph{,}1} }_{ \emph{h}_{ l\emph{,pq} },\emph{h}_{ l\emph{,p'q'} } }\big( \emph{t}, \tau \big) \nonumber \\ [0.25cm]
\hspace*{-0.25cm}& + & \hspace*{-0.25cm} \rho^{ \emph{DB}_{l\emph{,}2} }_{ \emph{h}_{l\emph{,pq}}, \emph{h}_{l\emph{,pq}} } \big( \emph{t}, \tau \big)
\end{eqnarray}

By applying the corresponding scatterer non-uniform distribution, and by following similar reasoning in [19,21], we can obtain the time-variant space CFs of the LoS, single-, and double-bounced rays, as outlined below. Specifically, by submitting (2) into (18), the time-variant space CF in the case of the LoS rays can be expressed as

\begin{small}
\begin{eqnarray}
\rho^{ \emph{LoS} }_{ \emph{h}_{1\emph{,pq}},\emph{h}_{1\emph{,p'q'}} } \big( \emph{t}, \tau \big) \hspace*{-0.25cm} &=& \hspace*{-0.25cm} \emph{K} e^{ j \frac{ 2\pi f_\emph{c} }{ \emph{c} } \sqrt{ \big( \emph{D} - \delta_\emph{Tx} \big)^2 + \big( \emph{v}_\emph{R} \emph{t} \big)^2 - 2\big( \emph{D} - \delta_\emph{Tx} \big) \emph{v}_\emph{R} \emph{t} \cos\big( \alpha^\emph{LoS}_\emph{R} - \gamma_\emph{R} \big) } } \nonumber \\ [0.2cm]
&\times& \hspace*{-0.25cm} e^{ j 2 \pi f_\emph{max} \tau \cos\big( \alpha^\emph{LoS}_\emph{R} - \gamma_\emph{R} \big) }
\end{eqnarray}
\end{small}

\hspace*{-0.5cm} where $\lambda$ denotes the wavelength. In submitting (3) into (18), the time-variant space CF in the case of the single-bounced rays $\emph{SB}_{1\emph{,}i}$ can be derived as

\begin{small}
\begin{eqnarray}
\rho^{\emph{SB}_{1\emph{,}i}}_{ \emph{h}_{1\emph{,pq}},\emph{h}_{1\emph{,p'q'}} } \big( \emph{t}, \tau \big) = \eta_{\emph{SB}_{1\emph{,}i}} \mathop {\lim } \limits_{{\emph{N}_{1\emph{,}i}} \to \infty } \sum_{ \emph{n}_{1\emph{,}i}=1 }^{\emph{N}_{1\emph{,}i}} \frac{1}{ \emph{N}_{1\emph{,}i} }  e^{ j \frac{ 2\pi f_\emph{c} }{ \emph{c} } \emph{A}^{(\emph{SB}_{1\emph{,}i})} + \emph{B}^{(\emph{SB}_{1\emph{,}i})} }
\end{eqnarray}
\end{small}

\hspace*{-0.6cm} where $\emph{A}^{(\emph{SB}_{1\emph{,}1})} = \big[ ( \emph{D} - \emph{Q}_\emph{q} \cos\theta_\emph{R} )^2 + ( \emph{v}_\emph{R} \emph{t} )^2 - 2 ( \emph{D} - \emph{Q}_\emph{q} \cos\theta_\emph{R} ) \times \emph{v}_\emph{R} \emph{t}  \cos( \alpha^{(\emph{n}_{1\emph{,}1})}_\emph{R} - \gamma_\emph{R} ) \big] - \delta_\emph{Tx} \cos\alpha^{(\emph{n}_{1\emph{,}1})}_\emph{T} \cos\beta^{(\emph{n}_{1\emph{,}1})}_\emph{T} - \delta_\emph{Ty} \times \sin\alpha^{(\emph{n}_{1\emph{,}1})}_\emph{T} \cos\beta^{(\emph{n}_{1\emph{,}1})}_\emph{T} - \delta_\emph{Tz} \sin\beta^{(\emph{n}_{1\emph{,}1})}_\emph{T} $, $\emph{B}^{(\emph{SB}_{1\emph{,}1})} = j 2 \pi f_\emph{max} \tau \cos( \alpha^{(\emph{n}_{1\emph{,}1})}_\emph{R} - \gamma_\emph{R} ) \cos\beta^{(\emph{n}_{1\emph{,}1})}_\emph{R} $, $\emph{A}^{(\emph{SB}_{1\emph{,}2})} = \big[ ( \emph{D} - \emph{Q}_\emph{p} \cos\theta_\emph{T} )^2 + ( \emph{v}_\emph{R} \emph{t} )^2 - 2 \big( \emph{D} - \emph{Q}_\emph{p} \big) \times \emph{v}_\emph{R} \emph{t} \cos( \alpha^{(\emph{n}_{1\emph{,}2})}_\emph{R} - \gamma_\emph{R} ) \big]^{1/2} - \delta_\emph{Rx} \cos\alpha^{(\emph{n}_{1\emph{,}2})}_\emph{R} \cos\beta^{(\emph{n}_{1\emph{,}2})}_\emph{R} - \delta_\emph{Ry} \sin\alpha^{(\emph{n}_{1\emph{,}2})}_\emph{R} \cos\beta^{(\emph{n}_{1\emph{,}2})}_\emph{R} - \delta_\emph{Rz} \sin\beta^{(\emph{n}_{1\emph{,}2})}_\emph{R} $, and $\emph{B}^{(\emph{SB}_{1\emph{,}2})} = j 2 \pi f_\emph{max} \tau \cos( \alpha^{(\emph{n}_{1\emph{,}2})}_\emph{R} - \gamma_\emph{R} ) \times \cos\beta^{(\emph{n}_{1\emph{,}2})}_\emph{R}$.

It is stated in [6] that $\sum_{\emph{n}_{1\emph{,}i}=1}^{\emph{N}_{1\emph{,}i}} 1/\emph{N}_{1\emph{,}i} = 1$ as $\emph{N}_{1\emph{,}i} \to \infty$. Thus, the total power of the $\emph{SB}_{1\emph{,}i}$ rays is proportional to $1/\emph{N}_{1\emph{,}i}$. This is equal to the infinitesimal power coming from the differential of the 3D angles, $\emph{d} \alpha^{(\emph{n}_{1\emph{,}i})}_\emph{R} \emph{d} \beta^{(\emph{n}_{1\emph{,}i})}_\emph{R}$, i.e., $1/\emph{N}_{1\emph{,}i} = p( \alpha^{(\emph{n}_{1\emph{,}i})}_\emph{R}, \beta^{(\emph{n}_{1\emph{,}i})}_\emph{R} )\emph{d} \alpha^{(\emph{n}_{1\emph{,}i})}_\emph{R} \emph{d} \beta^{(\emph{n}_{1\emph{,}i})}_\emph{R}$, where $p( \alpha^{(\emph{n}_{1\emph{,}i})}_\emph{R}, \beta^{(\emph{n}_{1\emph{,}i})}_\emph{R} )$ denotes the joint von Mises PDF in (17). Therefore, (22) can be rewritten as
\begin{eqnarray}
\rho^{\emph{SB}_{1\emph{,}i}}_{ \emph{h}_{1\emph{,pq}},\emph{h}_{1\emph{,p'q'}} } \big( \emph{t}, \tau \big) \hspace*{-0.25cm} & = & \hspace*{-0.25cm} \eta_{\emph{SB}_{1\emph{,}i}} \int_{ -\pi }^{\pi} \int_{-\pi}^{\pi} e^{ j \frac{ 2\pi f_\emph{c} }{ \emph{c} } \emph{A}^{(\emph{SB}_{1\emph{,}i})} + \emph{B}^{(\emph{SB}_{1\emph{,}i})}} \nonumber \\[0.2cm]
\hspace*{-0.25cm} & \times & \hspace*{-0.25cm} p \Big( \alpha^{ (\emph{n}_{1\emph{,}i}) }_\emph{R}, \beta^{ (\emph{n}_{1\emph{,}i}) }_\emph{R} \Big)  \emph{d} \alpha^{(\emph{n}_{1\emph{,}i})}_\emph{R} \emph{d} \beta^{(\emph{n}_{1\emph{,}i})}_\emph{R}
\end{eqnarray}

Similarly, submitting (4) into (18), the time-variant space CF in the case of the single-bounced rays $\emph{SB}_{l\emph{,}3}$ can be expressed as

\begin{eqnarray}
\rho^{\emph{SB}_{l\emph{,}3}}_{ \emph{h}_{l\emph{,pq}},\emph{h}_{l\emph{,p'q'}} } \big( \emph{t}, \tau \big) \hspace*{-0.25cm} & = & \hspace*{-0.25cm} \eta_{\emph{SB}_{l\emph{,}3}} \int_{ -\pi }^{\pi} \int_{-\pi}^{\pi} e^{ j \frac{ 2\pi f_\emph{c} }{ \emph{c} } \emph{A}^{(\emph{SB}_{l\emph{,}3})} + \emph{B}^{(\emph{SB}_{l\emph{,}3})}} \nonumber \\[0.2cm]
\hspace*{-0.25cm} & \times & \hspace*{-0.25cm} p \Big( \alpha^{ (\emph{n}_{l\emph{,}3}) }_\emph{R}, \beta^{ (\emph{n}_{l\emph{,}3}) }_\emph{R} \Big)  \emph{d} \alpha^{(\emph{n}_{l\emph{,}3})}_\emph{R} \emph{d} \beta^{(\emph{n}_{l\emph{,}3})}_\emph{R}
\end{eqnarray}

\hspace*{-0.55cm} where $\emph{A}^{(\emph{SB}_{l\emph{,}3})} = \big[ \xi^2_{\emph{p}\emph{n}_{l\emph{,}3}} ( \emph{t} ) \sin^2\beta^{(\emph{n}_{l\emph{,}3})}_\emph{T}  +  \xi^2_\emph{R} + ( \emph{v}_\emph{R} \emph{t} )^2 + 2 \emph{D} \times \xi_\emph{R} \cos( \gamma_\emph{R} + \alpha^{(\emph{n}_{l\emph{,}3})}_\emph{R} ) \big] - \delta_\emph{Tx} \cos\theta_\emph{T} \big[ \emph{R}_\emph{r} / \emph{D} \sin\psi_\emph{T} \sin\alpha^{(\emph{n}_{l\emph{,}3})}_\emph{T} + \cos\psi_\emph{T} \big] $ and $\emph{B}^{(\emph{SB}_{l\emph{,}3})} = j 2 \pi \tau \cos( \alpha^{(\emph{n}_{l\emph{,}3})}_\emph{R} - \gamma_\emph{R} ) \cos\beta^{(\emph{n}_{l\emph{,}3})}_\emph{R} $.

Submitting (6) into (18), the time-variant space CF in the case of the double-bounced rays \emph{DB} can be expressed as

\begin{eqnarray}
\rho^\emph{DB}_{ \emph{h}_{1\emph{,pq}},\emph{h}_{1\emph{,p'q'}} } \big( \emph{t}, \tau \big) \hspace*{-0.25cm} & = & \hspace*{-0.25cm} \eta_\emph{DB} \int_{-\pi}^{\pi} \int_{-\pi}^{\pi} e^{ j \frac{ 2\pi f_\emph{c} }{ \emph{c} } \emph{A}^{(\emph{DB})} + \emph{B}^{(\emph{DB})} } \nonumber \\[0.2cm]
\hspace*{-0.25cm} & \times & \hspace*{-0.25cm} p \Big( \alpha^{ (\emph{n}_{1\emph{,}2}) }_\emph{R}, \beta^{(\emph{n}_{1\emph{,}2})}_\emph{R} \Big)  \emph{d} \alpha^{(\emph{n}_{1\emph{,}2})}_\emph{R} \emph{d} \beta^{(\emph{n}_{1\emph{,}2})}_\emph{R}
\end{eqnarray}

\hspace*{-0.55cm} where $\emph{A}^{(\emph{DB}_{1\emph{,}2})} = \delta_\emph{Tx} \cos\alpha^{(\emph{n}_{1\emph{,}1})}_\emph{T} \cos\beta^{(\emph{n}_{1\emph{,}1})}_\emph{T} + \delta_\emph{Ty} \sin\alpha^{(\emph{n}_{1\emph{,}1})}_\emph{T} \times \cos\beta^{(\emph{n}_{1\emph{,}1})}_\emph{T} + \delta_\emph{Tz} \sin\beta^{(\emph{n}_{1\emph{,}1})}_\emph{T} + \delta_\emph{Rx} \cos\alpha^{(\emph{n}_{1\emph{,}2})}_\emph{R} \cos\beta^{(\emph{n}_{1\emph{,}2})}_\emph{R} + \delta_\emph{Ry}  \sin\alpha^{(\emph{n}_{1\emph{,}2})}_\emph{R} \cos\beta^{(\emph{n}_{1\emph{,}2})}_\emph{R} + \delta_\emph{Rz} \sin\beta^{(\emph{n}_{1\emph{,}2})}_\emph{R}$ and $\emph{B}^{(\emph{DB}_{1\emph{,}2})} = j 2 \pi \tau \times \cos( \alpha^{(\emph{n}_{1\emph{,}2})}_\emph{R} - \gamma_\emph{R} ) \cos\beta^{(\emph{n}_{1\emph{,}2})}_\emph{R} $.

Submitting (7) into (18), the time-variant space CF in the case of the double-bounced rays $\emph{DB}_{l\emph{,}1}$ can be derived as

\begin{eqnarray}
\rho^{\emph{DB}_{l\emph{,}1}}_{ \emph{h}_{l\emph{,pq}},\emph{h}_{l\emph{,p'q'}} } \big( \emph{t}, \tau \big) \hspace*{-0.25cm} & = & \hspace*{-0.25cm}  \eta_{\emph{DB}_{l\emph{,}1}} \int_{ -\pi }^{\pi} \int_{-\pi}^{\pi} e^{ j \frac{ 2\pi f_\emph{c} }{ \emph{c} } \emph{A}^{(\emph{DB}_{l\emph{,}1})} + \emph{B}^{(\emph{DB}_{l\emph{,}1})}} \nonumber \\[0.2cm]
\hspace*{-0.25cm} & \times & \hspace*{-0.25cm} p \Big( \alpha^{ (\emph{n}_{l\emph{,}3}) }_\emph{R}, \beta^{ (\emph{n}_{l\emph{,}3}) }_\emph{R} \Big)  \emph{d} \alpha^{(\emph{n}_{l\emph{,}3})}_\emph{R} \emph{d} \beta^{(\emph{n}_{l\emph{,}3})}_\emph{R}
\end{eqnarray}

\hspace*{-0.6cm} where $\emph{A}^{(\emph{DB}_{l\emph{,}1})} = \big[ \xi^2_{\emph{p}\emph{n}_{l\emph{,}3}} ( \emph{t} ) \sin^2\beta^{(\emph{n}_{l\emph{,}3})}_\emph{T}  +  \xi^2_\emph{R} + ( \emph{v}_\emph{R} \emph{t} )^2 + 2 \emph{D} \times \xi_\emph{R} \cos( \gamma_\emph{R} + \alpha^{(\emph{n}_{l\emph{,}3})}_\emph{R} ) \big] - \delta_\emph{Tx} \cos\alpha^{(\emph{n}_{1\emph{,}1})}_\emph{T} \cos\beta^{(\emph{n}_{1\emph{,}1})}_\emph{T} - \delta_\emph{Ty} \times \sin\alpha^{(\emph{n}_{1\emph{,}1})}_\emph{T}  \cos\beta^{ (\emph{n}_{1\emph{,}1}) }_\emph{T} - \delta_\emph{Tz} \sin\beta^{(\emph{n}_{1\emph{,}1})}_\emph{T}$ and $\emph{B}^{(\emph{DB}_{l\emph{,}1})} = j 2 \pi \times f_\emph{max} \tau \cos( \alpha^{(\emph{n}_{l\emph{,}3})}_\emph{R} - \gamma_\emph{R} ) \cos\beta^{(\emph{n}_{l\emph{,}3})}_\emph{R} $.

Submitting (8) into (18), the time-variant space CF in the case of the double-bounced rays $\emph{DB}_{l\emph{,}2}$ can be derived as

\begin{eqnarray}
\rho^{\emph{DB}_{l\emph{,}2}}_{ \emph{h}_{l\emph{,pq}},\emph{h}_{l\emph{,p'q'}} } \big( \emph{t}, \tau \big) \hspace*{-0.25cm} & = & \hspace*{-0.25cm}  \eta_{\emph{DB}_{l\emph{,}2}} \int_{ -\pi }^{\pi} \int_{-\pi}^{\pi} e^{ j \frac{ 2\pi f_\emph{c} }{ \emph{c} } \emph{A}^{(\emph{DB}_{l\emph{,}2})} + \emph{B}^{(\emph{DB}_{l\emph{,}2})}} \nonumber \\[0.2cm]
\hspace*{-0.25cm} & \times & \hspace*{-0.25cm} p \Big( \alpha^{ (\emph{n}_{1\emph{,}2}) }_\emph{R}, \beta^{ (\emph{n}_{1\emph{,}2}) }_\emph{R} \Big)  \emph{d} \alpha^{(\emph{n}_{1\emph{,}2})}_\emph{R} \emph{d} \beta^{(\emph{n}_{1\emph{,}2})}_\emph{R}
\end{eqnarray}

\hspace*{-0.475cm} where $\emph{A}^{(\emph{DB}_{l\emph{,}2})} = \delta_\emph{Tx} \emph{R}_\emph{r} / \emph{D} \sin\psi_\emph{T} \sin\alpha^{(\emph{n}_{l\emph{,}3})}_\emph{T} \cos\theta_\emph{T} + \delta_\emph{Tx} \cos\theta_\emph{T} \times \cos\psi_\emph{T} + \delta_\emph{Rx} \cos\alpha^{(\emph{n}_{1\emph{,}2})}_\emph{R} \cos\beta^{(\emph{n}_{1\emph{,}2})}_\emph{R} + \delta_\emph{Ry} \sin\alpha^{(\emph{n}_{1\emph{,}2})}_\emph{R} \cos\beta^{(\emph{n}_{1\emph{,}2})}_\emph{R} + \delta_\emph{Rz} \sin\beta^{(\emph{n}_{1\emph{,}2})}_\emph{R}$ and $\emph{B}^{(\emph{DB}_{l\emph{,}2})} = j 2 \pi f_\emph{max} \tau \cos( \alpha^{(\emph{n}_{1\emph{,}2})}_\emph{R} - \gamma_\emph{R} ) \times \cos\beta^{(\emph{n}_{1\emph{,}2})}_\emph{R}$.

From (21)-(27), we notice that the time-variant space CFs are related not only to the geometric model parameters, but also to the moving properties. By substituting (17) into (21)-(27), the time-variant space cross-functions for the LoS, single-, and double-bounced components can be respectively obtained. Furthermore, by setting $\emph{p} = \emph{p'}$ and $\emph{q} = \emph{q'}$, the time-variant space auto-correlation function (ACF) can be obtained [25]. On the other hand, we note that all above investigated statistical properties are time-variant on account of the non-WSS assumption of the proposed V2V channel model. Consequently, by applying the Fourier transformation of $\emph{h}_{l\emph{,pq}} ( \emph{t} )$, the time-variant frequency cross-correlation function of the proposed 3D non-stationary channel model can be derived as

\begin{small}
\begin{eqnarray}
\rho_{ \emph{h}_{ l\emph{,pq} }, \emph{h}_{l\emph{,p'q'}} } ( \emph{t}, \Delta f ) = \frac{ \emph{E} \Big[ \int_{ -\infty }^{ \infty }  \emph{h}_{l\emph{,pq}}( \emph{t}, \tau ) \hspace*{0.05cm} \emph{h}^*_{l\emph{,pq}} ( \emph{t}, \tau ) e^{ j 2\pi \Delta f \tau } \emph{d} \tau \Big] }{ \sqrt{ \emph{E} \Big[ \big| \emph{h}_{ l\emph{,pq} } ( \emph{t}, f ) \big|^2 \Big] \emph{E} \Big[ \big| \emph{h}_{ l\emph{,p'q'} } ( \emph{t}, f + \Delta f ) \big|^2 \Big] } }
\end{eqnarray}
\end{small}

Similar to the previous case, we substitute the corresponding channel response into (28), and the time-variant frequency CFs for the LoS, single-, and double-bounced propagation rays can be respectively derived. Nevertheless, the proposed channel model under the WSS assumption (i.e., $\emph{t}=0$) demonstrates that the channel statistics are not dependent on time \emph{t}. In this case, the proposed channel model tends to be a conventional F2M channel model.

Note that the above analysis is mainly for the flat communication environments, where the MR is far from the MT. However, when the MR is close to the MT, it is important to investigate the effect of ground reflection on the V2V channel statistics [35]. Here, we assume that there are $\emph{N}_\emph{g}$ effective scatterers uniformly existing on the ground in the azimuth plane. The heights of antennas mounted on the MT and MR are denoted as $\emph{H}_\emph{t}$ and $\emph{H}_\emph{r}$, respectively. The AAoA and EAoA of the waves scattered from the scatterer on the ground are denoted as $\alpha^{(\emph{n}_\emph{g})}_\emph{R}$ and $\beta^{(\emph{n}_\emph{g})}_\emph{R}$, respectively. The distances from the MT and MR to the scatterer on the ground are denoted as $\xi_{\emph{p}\emph{n}_\emph{g}}$ and $\xi_{\emph{q}\emph{n}_\emph{g}}$, respectively. The energy-related parameter for the NLoS rays of ground reflection is denoted $\eta_{\emph{SB}_\emph{g}}$. Therefore, the complex coefficient for the NLoS rays of ground reflection can be expressed as
\begin{eqnarray}
\emph{h}^{\emph{SB}_\emph{g}}_\emph{pq} ( \hspace*{-0.425cm} &\emph{t}& \hspace*{-0.425cm} ) = \sqrt{ \frac{ \eta_{\emph{SB}_\emph{g}} }{ \Omega +1} } \lim_{ \emph{N}_\emph{g} \to \infty} \sum_{ \emph{n}_\emph{g} = 1 }^{ \emph{N}_\emph{g} } \frac{1}{ \sqrt{ \emph{N}_\emph{g} } } e^{ - j2 \pi f_\emph{c} \big[ \xi_{\emph{p}\emph{n}_\emph{g}} + \xi_{\emph{q}\emph{n}_\emph{g}} \big] / \emph{c} } \nonumber \\ [0.2cm]
&\times& \hspace*{-0.25cm} e^{j 2\pi \emph{t} \times f_\emph{max} \cos\big( \alpha^{(\emph{n}_\emph{g})}_\emph{R} - \gamma_\emph{R} \big) \cos\beta^{(\emph{n}_\emph{g})}_\emph{R} }
\end{eqnarray}
Then, the corresponding space CF for the NLoS rays of ground reflection can be obtained in a similar method above, which is omitted here for brevity. In the model, we notice that the received signals scattered from the ground are more likely to be single-bounced, rather than double-bounced. Thus, the space CFs for the single-bounced rays of ground reflection should be considered.

\subsection{ Doppler PSD }

In the V2V channel, the signals can propagate from the MT to MR via different paths, each of which can involve reflection, diffraction, waveguiding, and so on. In addition to the fluctuations in the signal envelope and phase, the received signal frequency constantly varies as a result of the relative motion between the MT and MR. Here, let us define $\emph{S} (\gamma)$ as the Doppler spectrum of the proposed 3D V2V time-variant channel model. In this case, the received signals are formed by the single-bounced rays scattered from the scatterers located on the $l$th semi-ellipsoid, as well as the double-bounced rays caused by the scatterers from the combined single cylinder and the $l$th semi-ellipsoid. Moreover, it is assumed that the PDFs of the Doppler frequency at the MR are three independent random variables; thus, we can obtain the following characteristic functions as

\begin{eqnarray}
\rho^{\emph{SB}_{l\emph{,}3}}_{ \emph{h}_{l\emph{,pq}},\emph{h}_{l\emph{,p'q'}} } \big( \emph{t}, \omega \big) = \int_{ -\pi }^{\pi} \rho^{\emph{SB}_{l\emph{,}3}}_{ \emph{h}_{l\emph{,pq}}, \emph{h}_{l\emph{,p'q'}} } \big( \emph{t}, \Delta f \big)  e^{ j \omega \Delta f } \emph{d}  \Delta f
\end{eqnarray}
\begin{eqnarray}
\rho^{\emph{SB}_{l\emph{,}1}}_{ \emph{h}_{l\emph{,pq}},\emph{h}_{l\emph{,p'q'}} } \big( \emph{t}, \omega \big) = \int_{ -\pi }^{\pi} \rho^{\emph{DB}_{l\emph{,}1}}_{ \emph{h}_{l\emph{,pq}}, \emph{h}_{l\emph{,p'q'}} } \big( \emph{t}, \Delta f \big)  e^{ j \omega \Delta f } \emph{d}  \Delta f
\end{eqnarray}
\begin{eqnarray}
\rho^{\emph{SB}_{l\emph{,}2}}_{ \emph{h}_{l\emph{,pq}},\emph{h}_{l\emph{,p'q'}} } \big( \emph{t}, \omega \big) = \int_{ -\pi }^{\pi} \rho^{\emph{DB}_{l\emph{,}2}}_{ \emph{h}_{l\emph{,pq}}, \emph{h}_{l\emph{,p'q'}} } \big( \emph{t}, \Delta f \big)  e^{ j \omega \Delta f } \emph{d}  \Delta f
\end{eqnarray}

If we take (30)-(32) into the inverse Fourier transform formula, the PDF of the total Doppler frequency can be derived as

\begin{eqnarray}
\rho \big( \emph{t}, \Delta f \big) \hspace*{-0.25cm} & = & \hspace*{-0.25cm}  \frac{1}{ 2\pi } \int_{ -\pi }^{\pi} \rho^{\emph{SB}_{l\emph{,}3}}_{ \emph{h}_{l\emph{,pq}},\emph{h}_{l\emph{,p'q'}} } \big( \emph{t}, \omega \big) \rho^{\emph{SB}_{l\emph{,}1}}_{ \emph{h}_{l\emph{,pq}},\emph{h}_{l\emph{,p'q'}} } \big( \emph{t}, \omega \big) \nonumber \\[0.2cm]
\hspace*{-0.25cm} & \times & \hspace*{-0.25cm} \rho^{\emph{SB}_{l\emph{,}2}}_{ \emph{h}_{l\emph{,pq}},\emph{h}_{l\emph{,p'q'}} } \big( \emph{t}, \omega \big) e^{ j \omega \Delta f } \emph{d}  \Delta f
\end{eqnarray}

Subsequently, we define the Fourier transform of $\rho \big( \emph{t}, \Delta f \big)$ with respect to the variable \emph{t}. Then, we can obtain the function $\emph{S} (\gamma, \Delta f)$ as

\begin{eqnarray}
\emph{S} \big(\gamma, \Delta f \big) = \int_{ -\pi }^{\pi} \rho \big( \emph{t}, \Delta f \big) e^{ - j 2 \pi t \gamma } \emph{d}  \emph{t}
\end{eqnarray}

If we set $\Delta f = 0$, we can then obtain $\rho ( \emph{t}, 0 ) = \rho ( \emph{t} )$ and $\emph{S} (\gamma, 0 ) = \emph{S} (\gamma )$. Therefore, the equation in (34) can be rewritten as

\begin{eqnarray}
\emph{S} \big(\gamma \big) = \int_{ -\pi }^{\pi} \rho \big( \emph{t} \big) e^{ - j 2 \pi t \gamma } \emph{d}  \emph{t}
\end{eqnarray}

Thus far, the Doppler spectrum $\emph{S}(\gamma)$ can be obtained. Obviously, note that the proposed Doppler spectrum does not only depend on the proposed channel model parameters, but also on the non-stationary properties.

\section{ Numerical Results and Discussions }

\begin{table*}[htb]
\scriptsize
\caption{Channel related parameters used in the simulations}
\centering
\renewcommand\arraystretch{2}
\newcommand{\tabincell}[2]{\begin{tabular}{@{}#1@{}}#2\end{tabular}}
\begin{tabular}{ |c|c|c|c|c| }
\hline  & \tabincell{c}{Tap one highway \\ scenarios} & Tap one urban scenarios & \tabincell{c}{Tap two highway \\ scenarios} & Tap two urban scenarios \\
\hline All scenarios & \multicolumn{4}{|c|}{ $\emph{D} = 200$ m, $\emph{a}_1 = 120$ m, $\emph{a}_2 = 140$ m, $f_\emph{c} = 5.4$ GHz, $\emph{v}_\emph{R} = 54$ km/h, $\psi_\emph{T} = \theta_\emph{T} = \pi/3$, $\psi_\emph{R} = \theta_\emph{R} = \pi/3$. } \\
\hline Basic parameters & \tabincell{c}{$\emph{R}_\emph{t} = \emph{R}_\emph{r} = 40$ m, \\ $\emph{v}_\emph{R} = 25$ m/s, $f_\emph{max} = 433$ Hz} & \tabincell{c}{$\emph{R}_\emph{t} = \emph{R}_\emph{r} = 20$ m, \\ $\emph{v}_\emph{R} = 8.3$ m/s, $f_\emph{max} = 144$ Hz} & \tabincell{c}{$\emph{R}_\emph{t} = \emph{R}_\emph{r} = 40$ m, \\ $\emph{v}_\emph{R} = 25$ m/s, $f_\emph{max} = 433$ Hz} & \tabincell{c}{$\emph{R}_\emph{t} = \emph{R}_\emph{r} = 20$ m, \\ $\emph{v}_\emph{R} = 8.3$ m/s, $f_\emph{max} = 144$ Hz} \\
\hline Rician factor & $\Omega = 3.942$ & $\Omega = 1.062$ & $\Omega = 3.942$ & $\Omega = 1.062$ \\
\hline \tabincell{c}{Energy-related \\ parameters} & \tabincell{c}{ $\eta_{\emph{S}\emph{B}_{1\emph{,}1}} = 0.371$, $\eta_{\emph{S}\emph{B}_{1\emph{,}2}} = 0.212$, \\ $\eta_{\emph{S}\emph{B}_{1\emph{,}3}} = 0.402$, $\eta_\emph{DB} = 0.015$ } & \tabincell{c}{ $\eta_{\emph{S}\emph{B}_{1\emph{,}1}} = \eta_{\emph{S}\emph{B}_{1\emph{,}2}} = 0.142$, \\ $\eta_{\emph{S}\emph{B}_{1\emph{,}3}} = 0.085$, $\eta_\emph{DB} = 0.631$ } & \tabincell{c}{ $\eta_{\emph{S}\emph{B}_{2\emph{,}3}} = 0.724$, \\ $\eta_{ \emph{D}\emph{B}_{2\emph{,}1} } = \eta_{\emph{D}\emph{B}_{2\emph{,}2}} = 0.138$} & \tabincell{c}{ $\eta_{\emph{S}\emph{B}_{2\emph{,}3}} = 0.056$, \\ $\eta_{\emph{D}\emph{B}_{2\emph{,}1}} = \eta_{\emph{D}\emph{B}_{2\emph{,}2}} = 0.472$ }\\
\hline \tabincell{c}{Environment-related \\ parameters} & \tabincell{c}{$\emph{k}^{(1\emph{,}1)} = 8.9$, $\emph{k}^{(1\emph{,}2)} = 2.7$, \\ $\emph{k}^{(1\emph{,}3)} = 12.3$ } & \tabincell{c}{ $\emph{k}^{(1\emph{,}1)} = 0.55$, $\emph{k}^{(1\emph{,}2)} = 1.21$, \\ $\emph{k}^{(1\emph{,}3)} = 12.3$ } & \tabincell{c}{ $\emph{k}^{(2\emph{,}1)} = 8.9$, $\emph{k}^{(2\emph{,}2)} = 2.7$, \\ $\emph{k}^{(2\emph{,}3)} = 12.3$ } & \tabincell{c}{ $\emph{k}^{(2\emph{,}1)} = 0.55$, $\emph{k}^{(2\emph{,}2)} = 1.21$, \\ $\emph{k}^{(2\emph{,}3)} = 12.3$ } \\
\hline
\end{tabular}
\end{table*}

In this section, the statistical properties of the proposed 3D non-stationary wideband V2V channel model are evaluated and analyzed. The time slots for the stationary and non-stationary conditions are set $\emph{t}=0$ and $\emph{t}=2$ s, respectively. Here, in order to investigate the proposed channel statistics for different time delays, i.e., per-tap statistics, we define the semi-major dimensions for the first tap and second tap are respectively $\emph{a}_1 = 120$ m and $\emph{a}_2 = 140$ m, i.e., $\tau=2 (\emph{a}_2 - \emph{a}_1)/\emph{c} \approx 133$ ns $> 20$ ns. Unless otherwise specified, all the channel related parameters used in this section are listed in Table I. As mentioned before, the energy-related parameters for tap one and other taps should be equal to unity, i.e., $\sum_{i=1}^{3} \eta_{ \emph{SB}_{1\emph{,}i} } + \eta_\emph{DB} = 1$ and $\eta_{ \emph{SB}_{l\emph{,}3} } + \eta_{\emph{DB}_{l\emph{,}1} } + \eta_{\emph{DB}_{l\emph{,}2} } = 1$. Note that the energy-related parameters $\eta_{ \emph{SB}_{l\emph{,}1} }$, $\eta_{ \emph{SB}_{l\emph{,}2} }$, $\eta_{ \emph{SB}_{l\emph{,}3} }$, $\eta_\emph{DB}$, $\eta_{\emph{DB}_{l\emph{,}1} }$, and $\eta_{\emph{DB}_{l\emph{,}2} }$ are related to the scattered cases of NLoS rays, as in [36]. For example, the received scattered power in tap one highway scenarios is mainly from waves reflected by the stationary roadside environments. The moving vehicles represented by the scatterers located on the two cylinders are more likely to be single-bounced, rather than double-bounced [21,27]. This indicates that $ \eta_{\emph{SB}_{1\emph{,}3}} > \max \{ \eta_{\emph{SB}_{1\emph{,}1}}, \eta_{\emph{SB}_{1\emph{,}2}} \} > \eta_\emph{DB} $, i.e., $\eta_{\emph{SB}_{1\emph{,}3}} $ is normally larger than 0.4, $\eta_{\emph{SB}_{1\emph{,}1}}$ and $\eta_{\emph{SB}_{1\emph{,}2}}$ are normally both larger than 0.2 but smaller than 0.4, while $\eta_\emph{DB}$ is normally smaller than 0.1. For tap one urban scenarios, the received scattered power is mainly from the waves scattered from the two-cylinder model, i.e., $\eta_\emph{DB} > \max \{ \eta_{\emph{SB}_{1\emph{,}1}}, \eta_{\emph{SB}_{1\emph{,}2}}, \eta_{\emph{SB}_{1\emph{,}3}} \}$ (normally, $\eta_\emph{DB}$ is larger than 0.6, while $\eta_{\emph{SB}_{1\emph{,}1}}$, $\eta_{\emph{SB}_{1\emph{,}2}}$, and $\eta_{\emph{SB}_{1\emph{,}3}}$ are all smaller than 0.15). For tap two highway scenarios, the received scattered power is mainly from waves reflected by the stationary roadside environments described by the scatterers located on the semi-ellipsoid. Thus, $\eta_{\emph{SB}_{2\emph{,}3}} > \max \{ \eta_{\emph{DB}_{2\emph{,}1}}, \eta_{\emph{DB}_{2\emph{,}2}} \}$, i.e., $\eta_{\emph{SB}_{2\emph{,}3}}$ is normally larger than 0.7, while $\eta_{\emph{DB}_{2\emph{,}1}}$ and $\eta_{\emph{DB}_{2\emph{,}2}}$ are both smaller than 0.15). For tap two urban scenarios, the received scattered power is mainly from the double-bounced rays from the combined single cylinder and semi-ellipsoid models, i.e., $\min \{ \eta_{\emph{DB}_{2\emph{,}1}}, \eta_{\emph{DB}_{2\emph{,}2}} \} > \eta_{\emph{SB}_{2\emph{,}3}}$ (normally, $\eta_{\emph{SB}_{2\emph{,}3}}$ is smaller than 0.1, while $\eta_{\emph{DB}_{2\emph{,}1}}$ and $\eta_{\emph{DB}_{2\emph{,}2}}$ are both larger than 0.4). On the other hand, the environment-related parameters $\emph{k}^{(l\emph{,}1)}$, $\emph{k}^{(l\emph{,}2)}$, and $\emph{k}^{(l\emph{,}3)}$ are related to the distribution of scatterers. For example, higher values of $\emph{k}^{(l\emph{,}1)}$ and $\emph{k}^{(l\emph{,}2)}$ (i.e., normally both smaller than 10) result in the fewer moving vehicles/scatterers, i.e., the highway scenarios. In both the highway and urban scenarios, $\emph{k}^{(l\emph{,}3)}$ is large (i.e., normally larger than 10) as the scatterers reflected from roadside environments are normally concentrated. In addition, Ricean factor $\Omega$ is small (i.e., normally smaller than 1.5) in urban scenarios, as the LoS component does not have dominant power. However, $\Omega$ is large (i.e., normally larger than 3.5) in highway scenarios as fewer moving vehicles/obstacles (between the MT and MR) on the road result in the strong LoS propagation component.

\begin{figure}[t]
\begin{center}
\epsfig{file=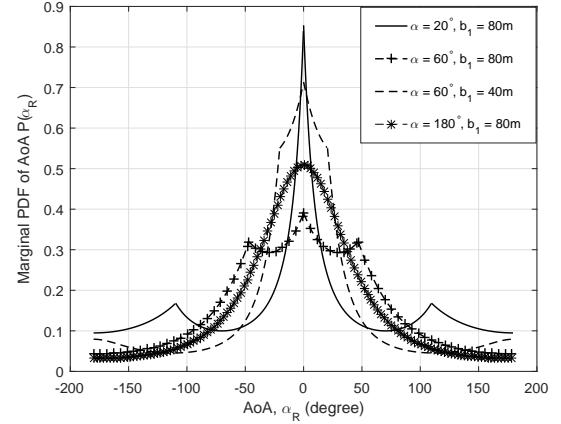, width=8cm} \caption{ Marginal PDF of the AoA statistics in the azimuth plane for the different channel parameter $\emph{b}_1$ and different beamwidths of the directional antenna at the MT.}
\end{center}
\end{figure}

\begin{figure}[t]
\begin{center}
\epsfig{file=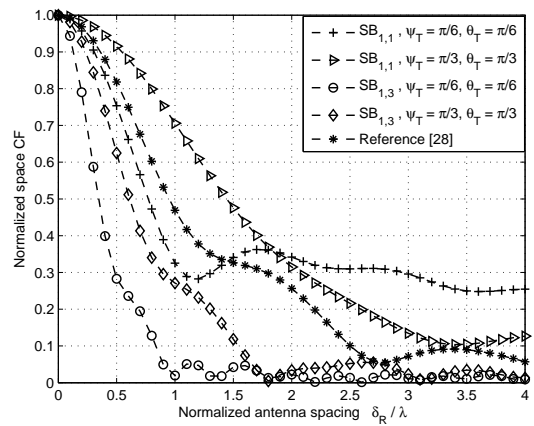, width=8cm} \caption{ Absolute values of time-variant space CFs of the single-bounced models for different transmit antenna angles in tap one highway scenarios. }
\end{center}
\end{figure}

Although the MT and MR in the proposed model are employed in ULA omni-directional antenna elements, the proposed model can also be used to analyze radiation patterns specific to the elements, which make the proposed geometric channel model irregularly shaped. Here, we assume that the transmitter emits the signal to the receiver in significantly small beamwidths, spanning the azimuth range of $[-\alpha, \alpha]$. It is stated in [16] that the AoA statistics of the multi-path components can be used to evaluate the performance of MIMO communication systems. Here, the marginal PDF of the AoA statistics corresponding to the road width $\emph{b}_1$ and the beamwidths of the directional antenna (i.e., $\alpha$) at the MT is shown in Fig. 5. It is apparent that, when the MT is employed with the directional antenna elements, the AoA PDFs in $0 \le \alpha^{(n_{l,3})}_\emph{R} \le \pi$ firstly decrease to a local value of AoA and then increase to a local maximum with a ``corner'', the AoA PDFs finally decrease sharply, depending upon the proposed geometric channel model, as seen in Figs. 1 and 2. A similar behavior can be seen in $-\pi \le \alpha^{(n_{l,3})}_\emph{R} \le 0$. By increasing the beamwidths $\alpha$ with more scatterers in the scattering region illuminated by the directional antenna, the PDFs firstly have higher values on both sides of the curves, and then gradually tend to be equal. It can also be noted that when the road width $\emph{b}_1$ increases from $40$ m to $80$ m, the values of the AoA PDFs increase sharply.

\begin{figure}[t]
\begin{center}
\epsfig{file=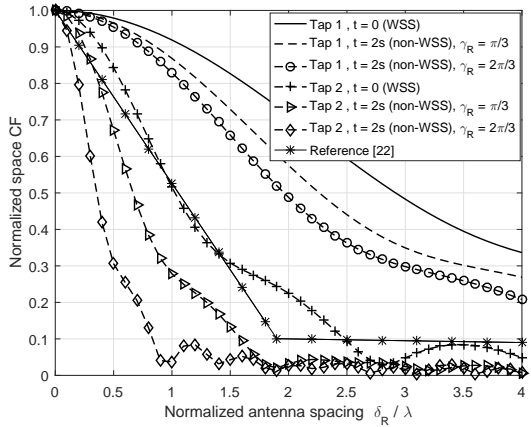, width=8cm} \caption{ Absolute values of the time-variant space CFs of the single-bounced semi-ellipsoid model for different taps of the proposed model in highway scenarios. }
\end{center}
\end{figure}

By adopting an MT antenna element spacing $\delta_\emph{T} = \lambda$, the absolute values of the time-variant space CF of the proposed V2V channel model are illustrated in Figs. 6, 7, and 8. By imposing $i=1$ and $3$ in (23), Fig. 6 shows the absolute values of time-variant space CFs of the single-bounced models (i.e., $\emph{SB}_{1\emph{,}1}$ and $\emph{SB}_{1\emph{,}3}$) for different transmit antenna azimuth angles $\psi_\emph{T}$ and elevation angle $\theta_\emph{T}$. It is obvious that the spatial correlation gradually decreases when the normalized antenna spacing $\emph{d} \cdot \lambda^{-1}$ increases. A similar behavior can be seen in [28]. Additionally, it is evident that the time-variant space CF decreases slowly as the transmit antenna angles (i.e., $\psi_\emph{T}$ and $\theta_\emph{T}$) decrease.

\begin{figure}[t]
\begin{center}
\epsfig{file=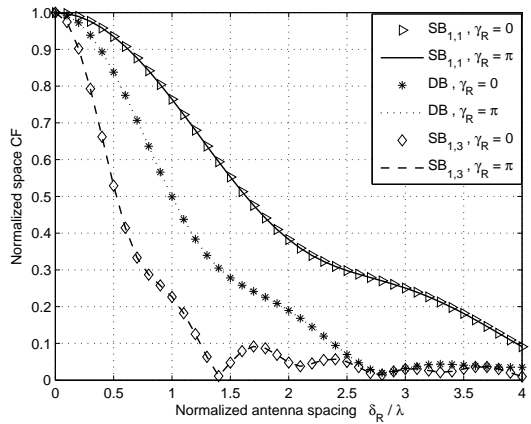, width=8cm} \caption{ Absolute values of the time-variant space CFs of the single- and double-bounced models for different relative moving directions in highway scenarios. }
\end{center}
\end{figure}
\begin{figure}[t]
\begin{center}
\epsfig{file=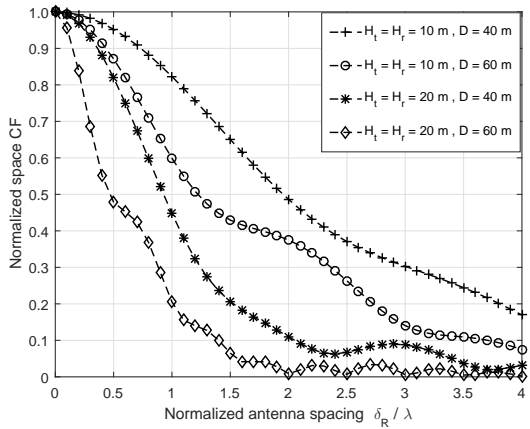, width=8cm} \caption{ Absolute values of the time-variant space CFs for the single-bounced rays of ground reflection for different antenna heights (i.e., $\emph{H}_\emph{t}$ and $\emph{H}_\emph{r}$) and different distances \emph{D} between the MT and MR. }
\end{center}
\end{figure}

Figs. 7 and 8 illustrate the absolute values of the time-variant space CFs for different channel conditions, i.e., WSS and non-WSS assumptions. By using (24), the absolute values of the time-variant space CFs of the first and second taps of the single-bounced semi-ellipsoid model (i.e., $\emph{SB}_{l\emph{,}3}$) for different taps and different relative moving properties (i.e., $\emph{t}$ and $\gamma_\emph{R}$) are shown in Fig. 7. In this figure, the higher correlation in the first tap is compared to the second tap because of the dominant LoS rays, which is in correspondence with the results in [22]. By using (25) and imposing $i=1$ and $3$ in (23), Fig. 8 illustrates the absolute values of the time-variant space CFs of the single- (i.e., $\emph{SB}_{1\emph{,}1}$ and $\emph{SB}_{1\emph{,}3}$) and double-bounced models (i.e., $\emph{DB}$) of the first tap in the WSS condition (i.e., $\emph{t}=0$). The figure shows that the relative moving directions (i.e., $\gamma_\emph{R}$) have no impact on the distribution of the time-variant space CFs when the proposed channel model is under the WSS assumption. It can be observed that the time-variant space CF of the single-bounced $\emph{SB}_{1\emph{,}3}$ is lower than that of the single-bounced $\emph{SB}_{1\emph{,}1}$. This is due to the fact that higher geometric path lengths result in lower correlation as mentioned in [27]. However, in the proposed model, the path length for $\emph{SB}_{1\emph{,}3}$ is obviously longer than the path length for $\emph{SB}_{1\emph{,}1}$.

Fig. 9 shows the time-variant space CFs for the single-bounced rays of ground reflection with respect to the different antenna heights (i.e., $\emph{H}_\emph{t}$ and $\emph{H}_\emph{r}$) and different distances \emph{D} between the MT and MR. From the figure, we can easily notice that when the heights of antennas mounted on the MT and MR increase from 10 m to 20 m, the space CFs decrease slowly, irrespective of the highway and urban environments. Additionally, the space CFs decrease gradually as the MR gets away from the MT. This is mainly due to the fact that higher geometric path lengths result in the lower correlation, as in Fig. 8.

\begin{figure}[t]
\begin{center}
\epsfig{file=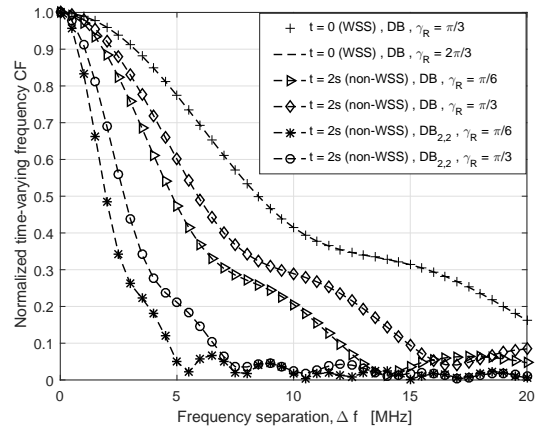, width=8cm} \caption{ Absolute values of the time-variant frequency CFs of the double-bounced models for different relative moving directions and different time instants in highway scenarios. }
\end{center}
\end{figure}

For V2V scenarios, it is important to analyze the impact of non-stationarity, including that of the relative moving directions (i.e., $\gamma_\emph{R}$) and moving time instants (i.e., $\emph{t}$), on the statistical properties of the proposed V2V channel model. Accordingly, by using (25) and (27), Fig. 10 shows the time-varying frequency CFs of the double-bounced models (i.e., \emph{DB} and $\emph{DB}_{2\emph{,}2}$) corresponding to the different relative moving directions and different moving time instants. It is clearly observed that, for the double-bounced \emph{DB} WSS model, regardless of what the relative moving directions are (i.e., $\gamma_\emph{R}=\pi/3$ or $2\pi/3$), the curves of the frequency CFs between them tend to be the same, which confirms the analysis in Fig. 6. Furthermore, it is evident that when the receiver's relative moving direction $\gamma_\emph{R}$ is $\pi/3$, the value of the time-variant frequency CF is relatively higher than that at $\gamma_\emph{R}=\pi/6$. This is because higher geometric path lengths result in lower correlation, whereas the path length for the path length at $\gamma_\emph{R}=\pi/3$ is obviously shorter than in the other cases [27]. Then, we observe that the frequency CF of the double-bounced $\emph{DB}_{2\emph{,}2}$ is lower than that of the double-bounced \emph{DB} in the proposed non-stationary V2V channel model. These results well align with those of previous work [12] and thus demonstrate the utility of our model.

\begin{figure}[t]
\begin{center}
\epsfig{file=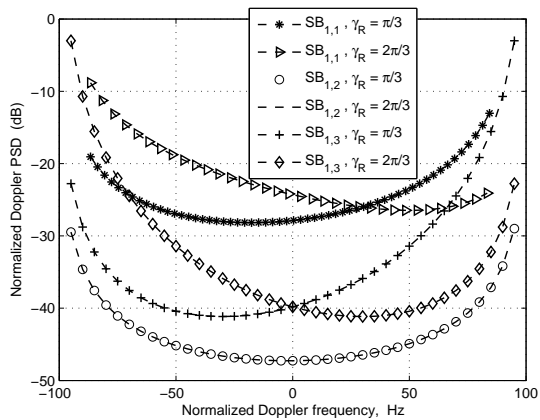, width=8cm} \caption{ Normalized Doppler PSDs of the proposed V2V channel model for different relative moving directions in highway scenarios. }
\end{center}
\end{figure}

\begin{figure}[t]
\begin{center}
\epsfig{file=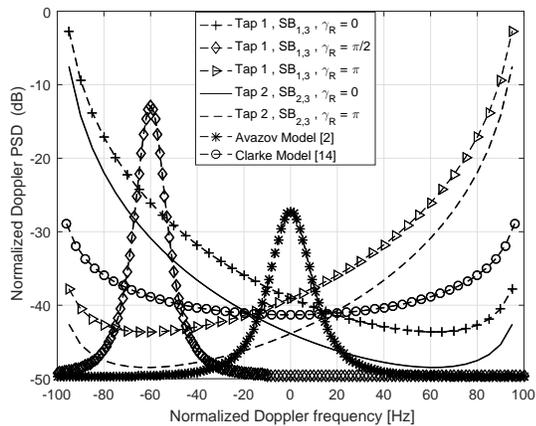, width=8cm} \caption{ Normalized Doppler PSDs of the single-bounced channel model for different taps and different relative moving directions in highway scenarios. }
\end{center}
\end{figure}

To understand the impact of the channel model parameters and non-stationary properties on Doppler PSDs given in (35) for the theoretical model, Fig. 11 shows the normalized Doppler PSDs of the proposed V2V channel model for different relative moving directions. It is observed that, for the direction of $\gamma_\emph{R} = \pi/3$, the Doppler PSD of the single-bounced $\emph{S}\emph{B}_{1\emph{,}1}$ is larger than that of the single-bounced $\emph{S}\emph{B}_{1\emph{,}3}$ because of the higher fading loss caused by the longer geometric path length. It is also evident that, for the waves that are single-bounced at the MR (i.e., $\emph{SB}_{1\emph{,}2}$), the relative moving direction has no impact on the distribution of the Doppler PSD. Moreover, this Doppler distribution tends to be a conventional U-shaped distribution, as shown in [10].

Fig. 12 shows the normalized Doppler PSDs of the single-bounced channel models (i.e., $\emph{SB}_{1\emph{,}3}$ and $\emph{SB}_{2\emph{,}3}$) for different taps and different relative moving directions (i.e., $\gamma_\emph{R}$). It is observed that the Doppler frequency gradually decreases with a decrease in the taps of the proposed channel model. It is also apparent that, for the MR movement perpendicular to the direct LoS rays (i.e., $\gamma_\emph{R} = \pi/2$), Doppler frequency in stationary channel model has a similar behavior to that of the results in [2] with a peak at zero. However, this is not necessary for the proposed non-stationary V2V channel model. We thus conclude that the Doppler spectrum in non-stationary V2V channels changes continually at different time instants when $\gamma_\emph{R}$ is set $\pi/2$, as reported in [25]. In addition, if we neglect the elevation angles around the receiver, the received signal comes from the single-bounced rays (i.e., $\emph{S}\emph{B}_{1\emph{,}2}$) caused by the scatterers uniformly located on a circle around the MR. Thus, the proposed Doppler PSD is given by the classic Clarke spectrum, which aligns with the results in [14].

Meanwhile, Fig. 13 illustrates the values of the impulse response of the proposed 3D model for different time delays. In the figure, time delay $\tau '$ can be defined as the ratio of the geometric path lengths and light velocity \emph{c}. The shortest and longest propagation delays of the proposed WSS model are respectively obtained as $\tau'_\emph{min} = \emph{D}/\emph{c}$ and $\tau'_\emph{max} \approx 2\emph{a}_l / \emph{c}$. Furthermore, it is evident that the impulse response gradually decreases with an increase in time delay $\tau'$, which agrees with the results in [27]. In addition, the channel response gradually decreases with an increase in the taps of the proposed channel model, which is in agreement with the theoretical analysis in Figs. 7 and 12. It is also apparent that the lower impulse channel is $\emph{R}_\emph{t}=20$ m compared to $\emph{R}_\emph{t}=40$ m because of the faster channel fading. The analysis above agrees with the results reported in [21], which can thus be fully utilized for the future design of wireless communication systems.

\begin{figure}[t]
\begin{center}
\epsfig{file=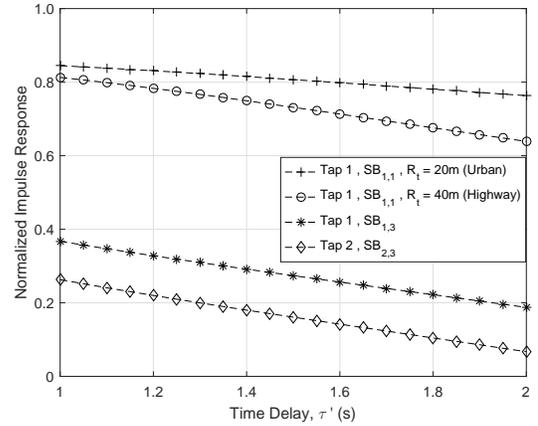, width=8cm} \caption{ Absolute value of the impulse response of the proposed single-bounced channel model for different taps and different V2V scenarios. }
\end{center}
\end{figure}

\section{ Conclusion }

In this paper, we have proposed a 3D wideband geometry-based channel model for V2V communication scenarios. The relative movement between the MT and MR results in the time-variant geometric statistics that make our model non-stationary. The proposed model adopts a two-cylinder model to depict moving vehicles (i.e., around the MT or MR), as well as multiple confocal semi-ellipsoid models to mimic stationary roadside environments. Based on experimental results, these channel statistics show different behaviors at different relative moving time instants, thereby demonstrating the capability of the proposed model in depicting a wide variety of V2V environments. It is additionally shown that the dominance of the LoS component results in a higher correlation in the first tap of the proposed channel model than in the second one. From the numerical results, we conclude that the time-variant space CF and frequency CF are significantly affected by the different taps of the proposed time-variant channel model, the relative moving times, and the directions between the MT and MR. Finally, it is shown that the proposed model closely agrees with the measured data, which validates the utility of our model.

\section{ Acknowledgements }

The authors would like to thank Professor Hikmet Sari, Department of Telecommunication and Information Engineering, Nanjing University of Posts and Telecommunications, China, for helping us complete this study successfully. The authors would also thank the anonymous reviewers for their constructive comments, which greatly helped improve this paper.

\end{document}